
\magnification\magstep 1
\parskip 5pt plus 1pt minus 0.5pt
\def\init{\tabskip 0pt}
\def\crr{\cr\noalign{\hrule}}
\def\euler{\gamma_{\scriptscriptstyle E}}
\def\frc#1#2{\displaystyle{#1\over #2}}
\def\Thetarms{\Theta_{\rm rms}}
\def\q{\kappa}
\def\d{{\rm d}}
\def\n{{\bf n}}
\def\p{\partial}
\def\r{{\bf r}}
\def\D{{\cal D}}
\def\O{{\cal O}}
\def\T{{\cal T}}
\def\Qabs{{\vert Q\vert}}
\def\Ai{\mathop{\rm Ai}}
\def\Bi{\mathop{\rm Bi}}
\def\Re{\mathop{\rm Re}}

\centerline{\bf ANISOTROPIC MULTIPLE SCATTERING IN DIFFUSIVE MEDIA}
\bigskip\null\medskip
\centerline{E. Amic,$^{(1)}$ J.M. Luck$^{(2)}$}
\medskip
\centerline{Service de Physique Th\'eorique, Centre d'\'Etudes de Saclay,}

\centerline{91191 Gif-sur-Yvette cedex, France,}
\bigskip
\centerline{Th.M. Nieuwenhuizen$^{(3)}$}
\medskip
\centerline{Van der Waals-Zeeman Institute, Valckenierstraat 65,}

\centerline{1018 XE Amsterdam, The Netherlands.}
\vfill
\noindent{\bf Abstract.}
The multiple scattering of scalar waves in diffusive media is investigated
by means of the radiative transfer equation.
This approach amounts to a resummation of the ladder diagrams
of the Born series; it does not rely on the diffusion approximation.
Quantitative predictions are obtained, concerning various
observables pertaining to optically thick slabs,
such as the mean angle-resolved reflected and transmitted intensities,
and the shape of the enhanced backscattering cone.
Special emphasis is put on the dependence of these quantities
on the anisotropy of the cross-section of the individual scatterers,
and on the internal reflections due to the optical index mismatch
at the boundaries of the sample.
The regime of very anisotropic scattering,
where the transport mean free path $\ell^*$ is much larger
than the scattering mean free path $\ell$, is studied in full detail.
For the first time the relevant Schwarzschild-Milne equation is solved
exactly in the absence of internal reflections,
and asymptotically in the regime of a large index mismatch.
An unexpected outcome concerns the angular width
of the enhanced backscattering cone,
which is predicted to scale as $\Delta\theta\sim\lambda/\sqrt{\ell\ell^*}$,
in contrast with the generally accepted $\lambda/\ell^*$ law,
derived within the diffusion approximation.
\vfill
\smallskip
\noindent To be submitted for publication to Physical Review E
\hfill SPhT/95/044
\smallskip
{\parskip 0pt
\noindent(1) e-mail: amic@amoco.saclay.cea.fr

\noindent(2) e-mail: luck@amoco.saclay.cea.fr

\noindent(3) e-mail: nieuwenh@phys.uva.nl
}
\eject

\noindent{\bf 1. INTRODUCTION}

The theory of multiple light scattering has been a classical subject of
interest for one century, which attracted the attention of many scientists,
including Lord Rayleigh, Schwarzschild, and Chandrasekhar.
Standard books are available, such as those
by Chandrasekhar [1], Ishimaru [2], and van de Hulst [3].
The discovery of weak-localization effects,
such as the celebrated enhanced backscattering cone [4],
yielded a revival of theoretical and experimental work in the area
of multiple scattering in disordered media.
Much progress has been done recently in the analysis
of speckle fluctuations [5],
in analogy with the conductance fluctuations observed in mesoscopic
electronic systems [6].

Laboratory experiments are often performed
either on solid TiO$_2$ (white paint) samples,
or on suspensions of polystyrene spheres or of TiO$_2$ grains in fluids.
In most cases, the wavelength $\lambda_0$ of light in the diffusive medium,
the scattering mean free path $\ell$, and the thickness $L$ of the sample
obey the inequalities $\lambda_0\ll\ell\ll L$.
Multiple scattering is also of interest in biophysics and medical physics,
in order to understand the transport of radiation
through human and animal tissues.
Besides light, all kinds of classical waves undergo multiple scattering
in media with a high enough level of disorder, i.e., of inhomogeneity.
Well-known examples are acoustic and seismic waves.
The propagation of electrons in disordered solids also pertains
to this area, since Quantum Mechanics also basically consists
in wave propagation.
Multiple scattering of waves in disordered media admits
the following three levels of theoretical description:

\noindent (i)
The {\it macroscopic} approach consists in an effective diffusion equation,
which describes the transport of the diffuse (incoherent)
intensity $I(\r,t)$ at point $\r$ at time $t$.
This approximation turns out to be very accurate in the bulk
of a turbid medium, and more generally on length scales much larger
than the mean free path $\ell$.
The diffusion approximation yields several interesting predictions,
among which we mention the $1/L$-decay of the total transmission
through an optically thick slab of thickness $L\gg\ell$,
the decay time of the transient response to an incident light pulse,
or the memory of a typical speckle pattern when the frequency of light
is varied.
This approximation also allows for a quantitative prediction
of the diffuse image of a small object in transmission [7].

\noindent (ii)
The {\it mesoscopic} approach, used by astrophysicists throughout
the classical era of the subject, is referred to as radiative transfer theory
[1--3].
This theory relies on the radiative transfer equation,
which is a local balance equation,
similar to the Boltzmann equation in kinetic theory,
for the diffuse intensity $I(\r,\n,t)$,
with $\n$ being the direction of propagation.
This approach leads in a natural way to distinguish between
the scattering mean free path $\ell$ and the transport mean free path $\ell^*$,
to be defined below.
The diffusion approach (i) is recovered
in the limit of length scales much larger than $\ell^*$.

\noindent (iii)
The {\it microscopic} approach consists in expanding the solution of the wave
equation in the disordered medium
in the form of a diagrammatic Born series.
In the regime $\ell\gg\lambda_0$ the leading diagrams can be identified,
in analogy with e.g. the theory of disordered superconductors [8].
For the diffuse intensity they are the celebrated ladder diagrams,
which are built up by pairing one retarded and one
advanced propagators following the same path through the disordered sample,
i.e., the same ordered sequence of scattering events.
This picture agrees with that behind the radiative transfer equation,
which is a classical transport equation for the intensity.
The ladder diagrams can be resummed and yield an integral equation
of the Bethe-Salpeter type for the diffuse intensity,
which we refer to as the Schwarzschild-Milne equation.
The radiative transfer approach (ii) is thus recovered in the
weak-disorder $(\ell\gg\lambda_0)$ regime.
A further step consists in including some of the subleading
diagrams, of relative order $\lambda_0/\ell\ll 1$,
which account for interference effects  between diffusive paths.
Among those contributions,
the class of maximally-crossed diagrams is of particular interest,
since it describes the aforementioned enhanced backscattering phenomenon [4].

To summarize this discussion, each of the descriptions mentioned above
represents a qualitative improvement with respect to the previous ones
(see e.g. ref. [9]).
In order to derive quantitative estimates of physical observables
in the regime $\ell\gg\lambda_0$ of experimental interest,
it is sufficient to consider radiative transfer theory.
The macroscopic approach (i) is clearly insufficient,
since we aim, among other things,
at a description of the crossover from free propagation to diffusive transport
which takes place in the {\it skin layers} of thickness of order $\ell$,
near the boundaries of the disordered medium.
This phenomenon is, by its very definition,
beyond the scope of the diffusion approach.
The radiative transfer approach (ii)
leads to study the Schwarzschild-Milne equation (2.24).
This equation has only been solved exactly in a limited number of cases.
Analytical results have been obtained for isotropic scattering
of scalar waves [1, 10--12],
and in the case where the scattering cross-section depends linearly
on the cosine of the scattering angle [1, 13].
The case of anisotropic scattering has essentially been investigated
by means of general formalism and by numerical methods [2, 3, 14].

For a finite sample, physical observables
such as the reflected or transmitted intensity in a given direction
depend on the particular realization of the sample under consideration.
Such quantities are indeed the results of intricate interference patterns
throughout the sample;
they only become self-averaging quantities in the limit
of a large enough sample.
This definition can be made precise by means of the dimensionless
conductance $g\sim{\cal N}\ell/L$,
related to the number ${\cal N}\sim{\cal A}/\lambda_0^2$ of open channels,
where ${\cal A}$ is the transverse area of the sample.
The self-averaging regime corresponds to $g\gg 1$.
The whole distribution of observables is therefore of physical interest,
as long as $g$ is not very large.
We mention a recent experiment [15],
where the third cumulant of the total transmission,
an effect of relative order $1/g^2$,
has been measured and compared to theoretical predictions.
This third cumulant is of the same order of magnitude
as the universal conductance fluctuations,
either in electronic [6] or in optical [5] systems.
In fact the full distribution of the total transmission
through an optically thick slab has been derived recently [16].
In the following we focus our attention on the mean values
of physical quantities.

The vector character of electromagnetic waves
also introduces its own intricacy.
In general four coupled integral equations
for two transfer functions have to be solved,
which are associated with the Stokes parameters
of the diffuse light in the medium.
These equations have been solved exactly in the case of Rayleigh scattering,
i.e., the regime where the size of the scatterers is much smaller
than the wavelength [1].
Among specific features pertaining to diffusive light propagation,
let us mention the dependence of the backscattering enhancement factor
on the polarization state of the outgoing diffuse radiation,
for a linearly polarized incident beam [17, 18],
or the progressive destruction of the backscattering peak
induced by a magnetic field,
due to the Faraday rotation in a magneto-optically active material [19, 20].
The latter effect has been recently observed
and interpreted quantitatively [21].

Furthermore,
in practical situations the optical index $n_0$ of the scattering medium is
often different from the index $n_1$ of the surrounding medium.
This index mismatch, measured by the ratio $m=n_0/n_1$,
causes reflections at the interfaces.
In the regime of a large index mismatch ($m\ll 1$ or $m\gg 1$),
the transmission across the interfaces is very small,
so that the light is reinjected many times in the diffusive medium.
As a consequence, the skin layers become very thick in this regime.
More generally, the diffusion approximation works better and better
as the index mismatch gets large.
Improvements of the diffusion equation have been proposed [22, 23],
which take internal reflections into account.
It is in fact possible to derive analytical expressions
for the reflected and transmitted intensities and other observables
in this regime.
This asymptotic analysis has been performed in ref. [12]
for isotropic scattering.
It will be generalized hereafter to the case of general anisotropic scattering.
This approach provides accurate results,
even for a moderate index mismatch $m$.

More generally, one of the main goals of this paper is to quantify
the dependence of physical quantities
on the anisotropy of the scattering mechanism.
It has been known for long from radiative transfer theory
that two length scales are involved in the case of general
anisotropic scattering: the scattering mean free path $\ell$,
namely the effective distance between two successive scattering events,
and the transport mean free path $\ell^*$,
namely the distance over which radiation looses memory of its direction.
Both mean free paths, to be defined more precisely in section 2,
depend on details of the scattering mechanism, such as the shape,
size, and dielectric constant of the scatterers.
In the regime of very anisotropic scattering, we have $\ell^*\gg\ell$.
The ratio $\tau^*=\ell^*/\ell\gg 1$ will be referred to as the
{\it anisotropy parameter}.
In some experimental situations $\tau^*$ can be of order ten or larger [24].
The regime of most physical interest is then $\lambda_0\ll\ell\ll\ell^*\ll L$.

The setup of this paper is as follows.
In section 2 we present some general formalism on the radiative transfer
equation, and we derive the associated Schwarzschild-Milne equation.
We show how solutions of the latter equation
yield predictions for quantities of interest,
such as the diffuse reflected and transmitted angle-resolved intensities,
and the shape of the enhanced backscattering cone.
Section 3 is devoted to the large index mismatch regime for general anisotropy.
In section 4 we derive a complete analytical solution of the problem
in the regime of very anisotropic scattering
(i.e., a cross-section strongly peaked in the forward direction),
in the absence of internal reflections.
The paper closes up with a discussion in section 5.

\medskip
\noindent{\bf 2. GENERALITIES ON ANISOTROPIC MULTIPLE SCATTERING}

Throughout the following
we restrict the analysis to multiple scattering of scalar waves
by scatterers located at uncorrelated random positions,
in the regime where the scattering mean free path $\ell$
is much larger than the wavelength $\lambda_0$ of radiation in the medium.
We have summarized some useful notations and definitions in Table 1.

As stated in section 1,
we put special emphasis on the dependence of physical quantities
on the anisotropy of the scattering mechanism.
It is known that,
after averaging over the random orientations of the individual scatterers,
the differential scattering cross-section
of arbitrary anisotropic scatterers can be written in the following form
[1, 2]
$$
\d\sigma(\n,\n')=(u/4\pi)^2 p(\Theta)\d\Omega'
.\eqno(2.1)
$$
In this formula $u$ is the scattering length,
$\n$ and $\n'$ are unit vectors in the incident and outgoing directions,
$\Theta$ is the angle between these directions, so that $\cos\Theta=\n.\n'$,
and $\d\Omega'$ is an element of solid angle around the direction $\n'$.

We assume that there is neither absorption,
nor inelastic scattering, i.e., that the albedo is unity
[the situation of non-conservative scattering will only
be considered in section 2.6].
The phase function $p(\Theta)$ then obeys the normalization condition
$$
\int{\d\Omega'\over 4\pi}p(\Theta)
=\int_{-1}^1{\d\cos\Theta\over 2}p(\Theta)=1
.\eqno(2.2)
$$
The total cross-section reads $\sigma=u^2/(4\pi)$,
and the {\it scattering mean free path} $\ell$ is given by
$$
\ell={1\over n\sigma}={4\pi\over nu^2}
,\eqno(2.3)
$$
where the density $n$ of scatterers is assumed to be small,
in such a way that we have $\ell\gg\lambda_0$.

In the particular case of isotropic scattering,
the phase function $p(\Theta)=1$ is constant.
In the general case of anisotropic scattering,
the phase function $p(\Theta)$ is a non-trivial function,
often peaked in the forward direction,
albeit in a more or less prominent way.
As recalled in section 1,
one has to distinguish between the scattering mean free path $\ell$,
which is the typical distance between two successive scattering events,
and the {\it transport mean free path} $\ell^*$,
which represents the distance over which radiation looses memory
of its direction.
The dimensionless ratio of both mean free paths,
$$
\tau^*={\ell^*\over\ell}={1\over 1-\langle\cos\Theta\rangle}
,\eqno(2.4)
$$
will be referred to as the {\it anisotropy parameter}.
We have introduced the notation
$$
\langle\cos\Theta\rangle=\int_{-1}^1{\d\cos\Theta\over 2}\cos\Theta\,p(\Theta)
.\eqno(2.5)
$$
We commonly have $\ell^*\ge\ell$,
since scattering usually preferentially takes place in the forward direction.
The regime of {\it very anisotropic scattering},
where $p(\Theta)$ is strongly peaked around the forward direction,
corresponds to $\ell^*\gg\ell$.
This regime will be investigated in detail in section 4.

\medskip
\noindent{\bf 2.1. General formalism and Schwarzschild-Milne equation}

In this section we present some general formalism
on anisotropic multiple scattering, extending thus the treatment
of the isotropic case presented in ref. [12].
Some of the results exposed below are already present in lecture notes
by one of us [9].

We begin with a derivation of the Schwarzschild-Milne equation.
This key equation of radiative transfer theory
will be the starting point of our analysis.
In the regime $\ell\gg\lambda_0$ under consideration,
in the absence of internal sources of radiation, and in stationary conditions,
the quantity of interest is the specific intensity $I(\r,\n)$
of radiation at the position $\r$, propagating in the direction $\n$.
The specific intensity obeys the time-independent radiative transfer equation,
that takes the following local form
$$
\ell\n.\nabla I(\r,\n)=\Gamma(\r,\n)-I(\r,\n)
.\eqno(2.6)
$$
The quantity
$$
\Gamma(\r,\n)=\int{\d\Omega'\over 4\pi}p(\n,\n')I(\r,\n')
\eqno(2.7)
$$
is commonly referred to as the source function.
We recall that the phase function $p(\n,\n')=p(\Theta)$
only depends on the scattering angle $\Theta$.

As recalled in section 1, the radiative transfer equation (2.6) [1--3]
can be considered as a mesoscopic balance equation for
the light intensity inside the diffusive medium,
somewhat analogous to the Boltzmann equation in the kinetic theory of gases.
It is equivalent to the Bethe-Salpeter equation,
obtained by resumming the ladder diagrams of the Born series
expansion of the intensity Green's function of the problem.
These diagrams are the dominant ones for $\ell\gg\lambda_0$,
i.e., to leading order in the density $n$ of scatterers.

We consider a sample of diffusive medium
in the form of a slab of thickness $L$,
limited by the two parallel planes $z=0$ and $z=L$.
The mean optical index $n_0$ of the slab can be different
from the index $n_1$ of the surrounding medium.
The index mismatch, measured by the ratio $m=n_0/n_1$,
generates internal reflections at the interfaces.
We introduce the optical depth $\tau=z/\ell$
of a point in the sample, and the optical thickness $b=L/\ell$ of the sample.
Most observables of interest can be derived
by integrating out the $\varphi$-dependence of the intensity $I(\r,\n)$,
i.e., by only considering quantities with cylindrical symmetry
around the normal to the slab.
The radiative transfer equation (2.6) then takes the form
$$
\mu{\d\over\d\tau}I(\tau,\mu)=\Gamma(\tau,\mu)-I(\tau,\mu)
,\eqno(2.8)
$$
or equivalently
$$
\mu{\d\over\d\tau}\big[I(\tau,\mu)e^{\tau/\mu}\big]
=\Gamma(\tau,\mu)e^{\tau/\mu}
.\eqno(2.9)
$$
The source function $\Gamma(\tau,\mu)$
is related to the specific intensity $I(\tau,\mu)$ through
$$
\Gamma(\tau,\mu)=\D\big[I(\tau,\mu)]
,\eqno(2.10)
$$
where we have introduced the integral operator $\D$
$$
\D[\Phi(\mu)]=\int_{-1}^1{\d\mu'\over 2}p_0(\mu,\mu')\Phi(\mu')
,\eqno(2.11)
$$
which involves the reduced phase function
$$
p_0(\mu,\mu')=\int_0^{2\pi}{\d\varphi'\over 2\pi}p(\mu,\varphi,\mu',\varphi')
.\eqno(2.12)
$$

The original phase function $p(\Theta)$
and the reduced one $p_0(\mu,\mu')$
are more simply related [1] through their expansions
in the Legendre polynomials
$$
\eqalign{
p(\Theta)&=\sum_{\ell\ge 0}(2\ell+1)\varpi_\ell P_\ell(\cos\Theta),\cr
p_0(\mu,\mu')&=\sum_{\ell\ge 0}(2\ell+1)\varpi_\ell P_\ell(\mu)P_\ell(\mu').
}\eqno(2.13)
$$

We mention for further reference that the Legendre polynomials
$\big\{P_\ell(\mu),\ell\ge 0\big\}$
form a complete set of orthogonal eigenfunctions
of the integral operator $\D$,
introduced in eq. (2.11), with eigenvalues $\varpi_\ell$, i.e.,
$$
\D[P_\ell(\mu)]=\varpi_\ell P_\ell(\mu)
.\eqno(2.14)
$$

The eigenvalue $\varpi_0=1$,
associated with the constant eigenfunction $P_0(\mu)=1$,
represents the unit albedo of the problem.
More explicitly,
$$
\int_{-1}^1{\d\mu'\over 2}p_0(\mu,\mu')=1
.\eqno(2.15)
$$
This identity yields the zero mode of the diffusion equation
in the long-distance limit.

The first non-trivial eigenvalue $\varpi_1$ is also of special interest.
It is indeed related to the anisotropy parameter $\tau^*$ of eq. (2.4).
Since $P_1(\mu)=\mu$, we have
$$
\varpi_1=\langle\cos\Theta\rangle=1-{1\over\tau^*}
.\eqno(2.16)
$$
More explicitly,
$$
\int_{-1}^1{\d\mu'\over 2}p_0(\mu,\mu')\mu'=\varpi_1\mu
=\langle\cos\Theta\rangle\mu
.\eqno(2.17)
$$

We also notice that all the eigenvalues of the operator $\D$
are trivial in the particular case of isotropic scattering,
since $\varpi_\ell=0$ for $\ell\ge 1$.

For simplicity, we consider first a half-space geometry $(b=\infty)$.
We assume that the limiting plane $\tau=0$ of the sample
is subjected to a cylindrically symmetric incident beam,
characterized by an angle of incidence $\theta_a$,
i.e., that the incident intensity
does not depend on the azimuthal angle $\varphi_a$.
This is automatically satisfied at normal incidence $(\theta_a=0)$.
Under these circumstances,
the inward intensity on the limiting plane $\tau=0^+$
contains both the normalized refracted incident beam,
coming in a direction defined by $\mu_a$,
and the intensity coming from the bulk of the medium,
after being reflected once at the interface.
Using the radiative transfer equation (2.8, 9), we thus get
$$
I(0^+,\mu,\mu_a)=2\delta(\mu-\mu_a)
+{R(\mu)\over\mu}\int_0^\infty\d\tau e^{-\tau/\mu}\Gamma(\tau,-\mu,\mu_a)
\quad(\mu>0)
,\eqno(2.18{\rm a})
$$
where the Fresnel reflection coefficient $R(\mu)$ is given in Table 1.
The only contribution to the outward intensity on this plane comes
from the light rays which have experienced at least one scattering event,
namely
$$
I(0^+,-\mu,\mu_a)
={1\over\mu}\int_0^\infty\d\tau e^{-\tau/\mu}\Gamma(\tau,-\mu,\mu_a)
\quad(\mu>0)
.\eqno(2.18{\rm b})
$$

The radiative transfer equation (2.8, 9),
together with the boundary conditions (2.18) at $\tau=0^+$,
can then be recast in the integral form
$$
I(\tau,\mu,\mu_a)
=2\delta(\mu-\mu_a)e^{-\tau/\mu_a}+(K*\Gamma)(\tau,\mu,\mu_a)
.\eqno(2.19)
$$
Here and throughout the following, the star denotes the convolution product
$$
(K*\Gamma)(\tau,\mu,\mu_a)
=\int_0^\infty\d\tau'\int_{-1}^1{\d\mu'\over 2}
K(\tau,\mu,\tau',\mu')\Gamma(\tau',\mu',\mu_a)
.\eqno(2.20)
$$
The kernel $K$ involved in this integral
can be split into two components: $K=K_B+K_L$.
The bulk kernel $K_B$ contains the contributions to the intensity at
depth $\tau$ arising from the scattering from either smaller or larger depths.
The layer kernel $K_L$ takes into account the intensity being scattered
at depth $\tau'$ in the direction of the wall,
then reflected there, and then scattered at depth $\tau$.
These kernels read explicitly
$$
\eqalign{
K_B(\tau,\mu,\tau',\mu')&=2\delta(\mu-\mu')\theta(\mu(\tau-\tau'))
{1\over\vert\mu\vert}e^{-(\tau-\tau')/\mu},\cr
K_L(\tau,\mu,\tau',\mu')
&=2\delta(\mu+\mu')\theta(\mu){R(\mu)\over\mu}e^{-(\tau+\tau')/\mu}.\cr
}\eqno(2.21)
$$

The final step consists in using eq. (2.10) in order to derive from eq. (2.19)
the following closed integral equation for the source function
$$
\Gamma(\tau,\mu,\mu_a)=p_0(\mu,\mu_a)e^{-\tau/\mu_a}+(M*\Gamma)(\tau,\mu,\mu_a)
,\eqno(2.22)
$$
which we refer to as the Schwarzschild-Milne equation of the problem.

The kernel $M$ has the following two components,
in analogy with the kernel $K$ from which it derives
$$
\eqalign{
M_B(\tau,\mu,\tau',\mu')&=\theta(\mu'(\tau-\tau')){p_0(\mu,\mu')\over
\vert\mu'\vert} e^{-(\tau-\tau')/\mu'},\cr
M_L(\tau,\mu,\tau',\mu')&=\theta(-\mu'){p_0(\mu,-\mu')\over
\vert\mu'\vert}R(-\mu')e^{-(\tau+\tau')/\vert\mu'\vert},
}\eqno(2.23)
$$
so that eq. (2.22) reads explicitly
$$
\eqalign{
\Gamma(\tau,\mu,\mu_a)&=p_0(\mu,\mu_a)e^{-\tau/\mu_a}\cr
&+\int_0^\tau\d\tau'\int_0^1{\d\mu'\over 2\mu'}p_0(\mu,\mu')
e^{-(\tau-\tau')/\mu'}\Gamma(\tau',\mu',\mu_a)\cr
&+\int_\tau^\infty\d\tau'\int_0^1{\d\mu'\over 2\mu'}p_0(\mu,-\mu')
e^{-(\tau'-\tau)/\mu'}\Gamma(\tau',-\mu',\mu_a)\cr
&+\int_0^\infty\d\tau'\int_0^1{\d\mu'\over 2\mu'}R(\mu')p_0(\mu,\mu')
e^{-(\tau+\tau')/\mu'}\Gamma(\tau',-\mu',\mu_a).\cr
}\eqno(2.24)
$$

In the case of isotropic scattering,
the phase function $p_0(\mu,\mu')=1$ is a constant,
and the source function $\Gamma(\tau,\mu,\mu_a)$ does not depend on $\mu$.
The Schwarzschild-Milne equation (2.24) thus takes a simpler form,
that has been extensively studied in ref. [12].
The rest of this section is devoted to an extension of the results
derived there to the general case of anisotropic scattering.

\medskip
\noindent{\bf 2.2. Solutions of Schwarzschild-Milne equation and sum rules}

In the general case of conservative anisotropic scattering,
the Schwarzschild-Milne equation (2.24) has a special solution
$\Gamma_S(\tau,\mu,\mu_a)$ which remains bounded as $\tau\to\infty$,
whereas the associated homogeneous equation has a linearly growing solution
$\Gamma_H(\tau,\mu)$.
More precisely, it can be checked by means of eq. (2.17)
that both source functions and the associated specific intensities
have the following asymptotic behavior for large depths $(\tau\gg 1)$,
up to exponentially small corrections
$$
\eqalign{
\Gamma_S(\tau,\mu,\mu_a)&\approx\tau_1(\mu_a),\cr
I_S(\tau,\mu,\mu_a)&\approx\tau_1(\mu_a),\cr
\Gamma_H(\tau,\mu)&\approx\tau-\mu(\tau^*-1)+\tau_0\tau^*,\cr
I_H(\tau,\mu)&\approx\tau-\mu\tau^*+\tau_0\tau^*.
}
\eqno(2.25)
$$

The quantities $\tau_0$ and $\tau_1(\mu_a)$ are unknown so far.
Let us anticipate that they play a central role in the following,
in the sense that they will bear the full non-trivial dependence of physical
quantities on the scattering mechanism.
These quantities also obey remarkable sum rules, to be derived below.

To do so, it is most convenient to introduce the Green's function
$G_S(\tau,\mu,\tau',\mu')$ of the problem, along the lines of ref. [12].
It is defined as the solution
which remains bounded as $\tau\to\infty$ of the equation
$$
G_S(\tau,\mu,\tau',\mu')
=p_0(\mu,\mu')\delta(\tau-\tau')+(M*G_S)(\tau,\mu,\tau',\mu')
.\eqno(2.26)
$$

The kernels $K$ and $M$
and the Green's function $G_S$ possess the symmetry properties
$$
\eqalign{
K(\tau,\mu,\tau',\mu')&=K(\tau',-\mu',\tau,-\mu),\cr
M(\tau,\mu,\tau',\mu')&=M(\tau',-\mu',\tau,-\mu),\cr
G_S(\tau,\mu,\tau',\mu')&=G_S(\tau',-\mu',\tau,-\mu),
}
\eqno(2.27)
$$
which merely express the time-reversal symmetry
of any sequence of scattering events.

As a consequence of eq. (2.22),
the special solution $\Gamma_S(\tau,\mu,\mu_a)$
can be expressed in terms of the Green's function as
$$
\Gamma_S(\tau,\mu,\mu_a)=\int_0^\infty\d\tau'
e^{-\tau'/\mu_a}G_S(\tau,\mu,\tau',\mu_a)
.\eqno(2.28)
$$

We also define for further reference the following bistatic coefficient
$$
\eqalign{
\gamma(\mu_a,\mu_b)
&=\int_0^\infty\d\tau e^{-\tau/\mu_b}\Gamma_S(\tau,-\mu_b,\mu_a)\cr
&=\int_0^\infty\d\tau e^{-\tau/\mu_b}\int_0^\infty\d\tau' e^{-\tau'/\mu_a}
G_S(\tau,-\mu_b,\tau',\mu_a).
}\eqno(2.29)
$$
The latter expression defines the bistatic coefficient
for any complex values of its arguments with $\Re\mu_a>0$, $\Re\mu_b>0$,
even outside the physical range $\mu_a,\mu_b\le 1$.
The symmetry $\gamma(\mu_a,\mu_b)=\gamma(\mu_b,\mu_a)$
is a consequence of the properties (2.27).
It is thus again due to time-reversal symmetry.

On the other hand,
a relationship between both solutions $\Gamma_H(\tau,\mu)$ and
$\Gamma_S(\tau,\mu,\mu_a)$
of the Schwarzschild-Milne equation can be derived as follows.
The Green's function $G_S(\tau,\mu,\tau',\mu')$
is clearly asymptotically proportional to the homogeneous solution
$\Gamma_H(\tau,\mu)$ when $\tau'$ goes to infinity, namely
$$
\lim_{\tau'\to\infty}G_S(\tau,\mu,\tau',\mu')={1\over D}\Gamma_H(\tau,\mu)
,\eqno(2.30)
$$
where the proportionality constant $D$ will be shown in a while
to be equal to the dimensionless diffusion coefficient (2.35).

As a consequence of eqs. (2.25, 28--30), we have
$$
\tau_1(\mu_a)=\lim_{\mu_b\to\infty}{\gamma(\mu_a,\mu_b)\over\mu_b}
={1\over D}\int_0^\infty\d\tau e^{-\tau/\mu_a}\Gamma_H(\tau,-\mu_a)
.\eqno(2.31)
$$

We now turn to the derivation of two groups of sum rules obeyed
by the quantities defined so far,
which are related to the so-called $F$ and $K$ integrals,
with the notations of Chandrasekhar [1].

The first group of two sum rules is a consequence of
the conservation of the flux in the $z$-direction
in a non-absorbing medium, given by the following $F$-integral
$$
F(\tau)=\int_{-1}^1{\d\mu\over 2}\mu I(\tau,\mu)
.\eqno(2.32)
$$
It can indeed be checked that eqs. (2.8, 15) yield $\d F/\d\tau=0$.

We investigate first the $F$-integral $F_S$ associated with the
special solution $\Gamma_S(\tau,\mu,\mu_a)$.
Considering the $\tau\to\infty$ limit determines $F_S=0$,
whereas considering the $\tau\to 0$ limit yields the sum rule
$$
\int_0^1{\d\mu\over 2}T(\mu)\gamma(\mu,\mu_a)=\mu_a
,\eqno(2.33)
$$
where the transmission coefficient $T(\mu)=1-R(\mu)$ is given in Table 1.
The $\mu_a\to\infty$ limit of eq. (2.33),
together with eq. (2.31), yields another sum rule, namely
$$
\int_0^1{\d\mu\over 2}T(\mu)\tau_1(\mu)=1
.\eqno(2.34)
$$
We can evaluate in a similar way the $F$-integral $F_H$ associated with the
homogeneous solution $\Gamma_H(\tau,\mu)$.
This yields no independent sum rule, but leads to the identification
of $D$ with the dimensionless diffusion coefficient
$$
D=\tau^*/3
.\eqno(2.35)
$$
The diffusion coefficient indeed reads $D_{\rm phys}=c\ell D=c\ell^*/3$
in physical units [1--3].
The transport velocity indeed coincides with the velocity of light
in vacuum $c$, to leading order in the regime $\ell\gg\lambda_0$.

Besides the sum rules (2.33, 34),
which were already given in ref. [12] for isotropic scattering,
the radiative transfer equation also admits another group of two sum rules,
which are novel in this context, and whose physical origin is less evident.
Consider the so-called $K$-integral,
again with the notations of Chandrasekhar [1]
$$
K(\tau)=\int_{-1}^1{\d\mu\over 2}\mu^2I(\tau,\mu)
.\eqno(2.36)
$$
It can be checked that eqs. (2.8, 16, 17) yield $\d K/\d\tau=-F/\tau^*$,
whence $K(\tau)=-F\tau/\tau^*+K_0$, with $K_0$ being independent of $\tau$.
By considering the $K$-integrals
associated with the special solution $\Gamma_S(\tau,\mu,\mu_a)$
and with the homogeneous solution $\Gamma_H(\tau,\mu)$,
we obtain after some algebra the following sum rules
$$
\eqalignno{
\int_0^1{\d\mu\over 2}\big(1+R(\mu)\big)\mu\gamma(\mu,\mu_a)
&={\tau_1(\mu_a)\over 3}-\mu_a^2,&(2.37{\rm a})\cr
\int_0^1{\d\mu\over 2}\big(1+R(\mu)\big)\mu\tau_1(\mu)
&=\tau_0.&(2.37{\rm b})\cr
}
$$

\medskip
\noindent{\bf 2.3. Diffuse reflected intensity}

The evaluation of the angle-resolved diffuse reflected intensity
by means of the general formalism exposed above
closely follows the lines of ref. [12].
We consider a half-space geometry,
and we assume that the limiting plane of the sample
is subjected to a cylindrically symmetric incident beam,
characterized by an angle of incidence $\theta_a$.
This technical assumption includes the case
of a plane wave at normal incidence $(\theta_a=0)$.
Under these circumstances,
the diffuse reflected intensity per solid angle $\d\Omega_b$ reads
$$
{\d R(a\to b)\over\d\Omega_b}=A^R(\theta_a,\theta_b)
={\cos\theta_a\over 4\pi m^2}{T_a T_b\over\mu_a\mu_b}\gamma(\mu_a,\mu_b)
,\eqno(2.38)
$$
where we have used some of the definitions of Table 1:
$m=n_0/n_1$ is the index mismatch,
and $T_a=T(\mu_a)$ and $T_b=T(\mu_b)$ are the transmission coefficients
in the incident and outgoing directions, respectively.

The most non-trivial part of the above result (2.38)
consists in the bistatic coefficient $\gamma(\mu_a,\mu_b)$,
whose definition and general properties have been exposed in section 2.2.
It will be evaluated in the large index mismatch regime in section 3,
and in the very anisotropic regime in section 4.

\medskip
\noindent{\bf 2.4. Diffuse transmitted intensity}

In this section we consider the angle-resolved mean transmission
of an optically thick slab,
of thickness $L=b\ell$, with $b\gg 1$ being large but finite.
A generalization of the reasoning of section 2.1 allows us to write
down the following Schwarzschild-Milne equation in this geometry
$$
\eqalign{
\Gamma_b(\tau,\mu,\mu_a)&=p_0(\mu,\mu_a)e^{-\tau/\mu_a}\cr
&+\int_0^\tau\d\tau'\int_0^1{\d\mu'\over 2\mu'}p_0(\mu,\mu')
e^{-(\tau-\tau')/\mu'}\Gamma_b(\tau',\mu',\mu_a)\cr
&+\int_\tau^b\d\tau'\int_0^1{\d\mu'\over 2\mu'}p_0(\mu,-\mu')
e^{-(\tau'-\tau)/\mu'}\Gamma_b(\tau',-\mu',\mu_a)\cr
&+\int_0^b\d\tau'\int_0^1{\d\mu'\over 2\mu'}R(\mu')p_0(\mu,\mu')
e^{-(\tau+\tau')/\mu'}\Gamma_b(\tau',-\mu',\mu_a)\cr
&+\int_0^b\d\tau'\int_0^1{\d\mu'\over 2\mu'}R(-\mu')p_0(\mu,-\mu')
e^{-(2b-\tau-\tau')/\mu'}\Gamma_b(\tau',\mu',\mu_a).
}\eqno(2.39)
$$

The solution of this equation for a thick slab
can be constructed from both solutions $\Gamma_S$ and $\Gamma_H$
of the half-space geometry, by means of a matching procedure,
along the lines of ref. [12].
Using the asymptotic forms (2.25), we obtain
$$
\Gamma_b(\tau,\mu,\mu_a)\approx\left\{\matrix{
\Gamma_S(\tau,\mu,\mu_a)
-\displaystyle{\tau_1(\mu_a)\over b+2\tau_0\tau^*}
\Gamma_H(\tau,\mu)\hfill&(\tau\hbox{ finite, } b-\tau\gg 1),\hfill\cr\null\cr
\displaystyle{\tau_1(\mu_a)\over
b+2\tau_0\tau^*}\Gamma_H(b-\tau,-\mu)\hfill&(b-\tau\hbox{ finite, }
\tau\gg 1).\hfill\cr}\right.
\eqno(2.40)
$$
Both expressions lead to a linear (diffusive) behavior
[3, 12] in the bulk of the sample ($\tau\gg 1$, $b-\tau\gg 1$), namely
$$
\Gamma_b(\tau,\mu,\mu_a)=\displaystyle{\tau_1(\mu_a)\over
b+2\tau_0\tau^*}\big[b-\tau+\tau_0\tau^*+\mu(\tau^*-1)\big]
+\O\big(e^{-\tau},e^{-(b-\tau)}\big)
.\eqno(2.41)
$$

The derivation of the diffuse transmitted intensity
per solid angle element $\d\Omega_b$
again closely follows the lines of ref. [12].
We obtain
$$
{\d T(a\to b)\over\d\Omega_b}
={\tau^*\over b+2\tau_0\tau^*}A^T(\theta_a,\theta_b)
,\eqno(2.42)
$$
with
$$
A^T(\theta_a,\theta_b)={\cos{\theta_a}\over 12\pi m^2}
{T_aT_b\over\mu_a\mu_b}\tau_1(\mu_a)\tau_1(\mu_b)
,\eqno(2.43)
$$
where we have again used some of the definitions of Table 1:
$m=n_0/n_1$ is the index mismatch,
and $T_a=T(\mu_a)$ and $T_b=T(\mu_b)$ are the transmission coefficients
in the incident and outgoing directions, respectively.

The most non-trivial part of the above result consists in the function
$\tau_1(\mu)$,
whose definition and general properties have been exposed in section 2.2.
It will be evaluated in the large index mismatch regime in section 3,
and in the very anisotropic regime in section 4.

The result (2.42) shows that the effective thickness of the sample
is $(b+2\tau_0\tau^*)\ell=L+2\tau_0\ell^*$.
In other words, $z_0=\tau_0\ell^*$ represents the thickness of a skin layer.
This quantity is also referred to as the injection depth,
or the extrapolation length of the problem.
Our result thus shows on a firm basis that the relevant length scale of the
problem in the diffusive regime is the transport mean free path $\ell^*$,
rather than the scattering mean free path $\ell$.
Let us anticipate that such a simple picture is not always correct.
Our expressions of the width of the backscattering cone
and of the absorption length will indeed involve
both mean free paths through the combination $\sqrt{\ell\ell^*}$.

\medskip
\noindent{\bf 2.5. Enhanced backscattering cone}

The general formalism exposed above can be extended
to the study of the celebrated enhanced backscattering phenomenon,
which takes place in a narrow cone
in the vicinity of the exact backscattering direction [4].
This phenomenon is due to the constructive interference between any
path in the medium and its time-reversed counterpart.
It is well-known that, for isotropic scattering,
the width of the cone is of order $\Delta\theta\sim\lambda_1/\ell\ll 1$.
One of the goals of this section is to derive a quantitative estimate
of the width of the cone, with emphasis on its dependence on the anisotropy
of the scattering mechanism.

As recalled in section 1,
the form of the backscattering cone can be predicted
by resumming the so-called maximally-crossed diagrams.
We restrict the analysis to normal incidence
$(\theta_a=0)$, and to the geometry of a half-space diffusive medium.
We introduce a dimensionless transverse wavevector
$$
Q=q\ell=k_0\ell\theta'=k_1\ell\theta
,\eqno(2.44)
$$
where $\theta'$ and $\theta$ are the incidence angles of the outgoing
radiation, and $k_0$ and $k_1$ are its wavenumbers,
respectively inside and outside the diffusive medium.

According to ref. [12], the reflected intensity in the vicinity
of the backscattering direction,
i.e., for $\theta\ll 1$, $k_1\ell\gg 1$, and $Q$ fixed, takes the form
$$
A^C(Q)\approx{T(1)^2\over 4\pi m^2}
\big[\gamma(1,1)+\gamma_C(Q)-p_0(1,-1)/2\big]
.\eqno(2.45)
$$

In this expression, $\gamma(1,1)$ represents the background reflected
intensity in the normal direction, in agreement with eq. (2.38).
The second term $\gamma_C(Q)$ is the contribution of the interference
between the sequences of any number $(n\ge 1)$ of scattering events
and their time-reversed counterparts.
The subtracted third term is the contribution of the single-scattering events
$(n=1)$, which are invariant under time inversion,
and should not be counted twice.
We define the enhancement factor
$$
B(Q)={A^C(Q)\over A^R(0,0)}
=1+{\gamma_C(Q)-p_0(1,-1)/2\over\gamma(1,1)}
.\eqno(2.46)
$$
It will turn out that the peak value of the interference contribution
coincides with the background term, i.e., $\gamma_C(0)=\gamma(1,1)$,
so that the enhancement factor at the top of the cone, namely
$$
B(0)=2-{p_0(1,-1)\over 2\gamma(1,1)}
,\eqno(2.47)
$$
nearly equals two,
up to the small contribution of single-scattering events.

 From a more quantitative viewpoint [9, 12], we have
$$
\gamma_C(Q)=\int_0^\infty\d\tau e^{-\tau}\Gamma_C(Q,\tau,-1)
,\eqno(2.48)
$$
where the function $\Gamma_C(Q,\tau,\mu)$ is a solution of
a $Q$-dependent Schwarzschild-Milne equation of the form
$$
\Gamma_C(Q,\tau,\mu)=p_0(\mu,1)e^{-\tau}+(M*\Gamma_C)(Q,\tau,\mu)
,\eqno(2.49)
$$
with a $Q$-dependent kernel $M=M_B+M_L$, in analogy with eq. (2.23), namely
$$
\eqalign{
M_B(Q,\tau,\mu,\tau',\mu')&=\theta(\mu'(\tau-\tau')){p_0(\mu,\mu')\over
\vert\mu'\vert}e^{-(\tau-\tau')/\mu'}
J_0\left(Q\vert\tau-\tau'\vert\sqrt{\mu'^{-2}-1}\right),\cr
M_L(Q,\tau,\mu,\tau',\mu')&=\theta(-\mu'){p_0(\mu,-\mu')\over
\vert\mu'\vert}R(-\mu')e^{-(\tau+\tau')/\vert\mu'\vert}
J_0\left(Q(\tau+\tau')\sqrt{\mu'^{-2}-1}\right),
}\eqno(2.50)
$$
where $J_0$ is the Bessel function.

The shape of the top of the backscattering cone,
described by the small-$Q$ behavior of $\gamma_C(Q)$,
is expected to be universal.
It is indeed due to the contribution of long paths in the diffusive medium,
along which the radiation undergoes many scattering events.
On the contrary, the tails of the cone,
corresponding to a large reduced wavevector $Q$,
only involve short sequences of two, three, etc. scattering events,
and are therefore expected to depend on the details
of the scattering mechanism.
This is already apparent in the subtracted term in eqs. (2.46, 47),
which involves the single-scattering cross-section
in the direction of exact backscattering.
We shall come back to this point in section 5.

The universal small-$Q$ behavior of the backscattering cone can be
determined analytically as follows.
Deep in the bulk of the medium, i.e., for $\tau\gg 1$,
the integral Schwarzschild-Milne equation (2.49)
can be approximated by a differential equation,
obtained by expanding the function
$\Gamma_C(Q,\tau,\mu)$ in powers of $(\tau'-\tau)$.
Keeping only the first two derivatives,
we obtain the following diffusion-like equation
$$
\eqalign{
\Gamma_C(Q,\tau,\mu)=\D&\left[\Gamma_C(Q,\tau,\mu)
-\mu{\d\over\d\tau}\Gamma_C(Q,\tau,\mu)\right.\cr
&\left.
+\mu^2{\d^2\over\d\tau^2}\Gamma_C(Q,\tau,\mu)
-{Q^2\over 2}(1-\mu^2)\Gamma_C(Q,\tau,\mu)
+\cdots\right].
}
\eqno(2.51)
$$
The asymptotic behavior of the source function $\Gamma_C(Q,\tau,\mu)$
for $\tau\gg 1$ and $Q\ll 1$ can be derived by looking for
a solution to eq. (2.51) in the form
of the following expansion in Legendre polynomials
$$
\Gamma_C(Q,\tau,\mu)=e^{-s_0\tau}\sum_{\ell\ge 0}c_\ell P_\ell(\mu)
.\eqno(2.52)
$$
Along the lines of Appendix A, and making use of eqs. (2.14) and (A.4),
we obtain the following recursion relations for the coefficients $c_\ell$,
involving the coefficients $\varpi_\ell$
of the expansion (2.13) of the phase function
$$
\eqalign{
{c_\ell\over\varpi_\ell}&=\left(1-{Q^2\over2}\right)c_\ell
+s_0\left({\ell\over2\ell-1}c_{\ell-1}+{\ell+1\over2\ell+3}c_{\ell+1}\right)\cr
&+\left(s_0^2+{Q^2\over2}\right)
\left({\ell(\ell-1)\over(2\ell-1)(2\ell-3)}c_{\ell-2}
+{2\ell^2+2\ell-1\over(2\ell-1)(2\ell+3)}c_\ell
+{(\ell+1)(\ell+2)\over(2\ell+3)(2\ell+5)}c_{\ell+2}\right).
}
\eqno(2.53)
$$

It is clear from the structure of these relations that,
when $Q$ is small, $s_0$ scales as $Q$,
and the coefficients $c_\ell\sim Q^\ell$ decay rapidly.
Keeping this hierarchy in mind, and using $\varpi_0=1$,
we expand the relations (2.53) for $\ell=0$ and $\ell=1$ as follows
$$
0={s_0^2-Q^2\over 3}c_0+{s_0\over 3}c_1+\cdots,\quad
\left({1\over\varpi_1}-1\right)c_1=s_0c_0+\cdots
\eqno(2.54)
$$
Using eq. (2.16), we obtain
$$
s_0\approx{\Qabs\over\sqrt{\tau^*}},
\quad {c_1\over c_0}\approx(\tau^*-1){\Qabs\over\sqrt{\tau^*}}\quad(Q\ll 1)
.\eqno(2.55)
$$

The next step also follows the lines of ref. [12].
The small-$Q$ expansion of the $Q$-dependent kernels (2.50)
only contains even powers of $Q$.
As a consequence,
the term proportional to $\Qabs$ in $\Gamma_C(Q,\tau,\mu)$
is proportional to the homogeneous solution $\Gamma_H(\tau,\mu)$
of the ordinary $(Q=0)$ Schwarzschild-Milne equation.
Putting everything together, we are left with the following estimates
of the source function $\Gamma_C(Q,\tau,\mu)$.
For $Q\ll 1$ and fixed $\tau$ we have
$$
\Gamma_C(Q,\tau,\mu)=\Gamma_S(\tau,\mu,1)
-\tau_1(1){\Qabs\over\sqrt{\tau^*}}\Gamma_H(\tau,\mu)+\O(Q^2)
,\eqno(2.56)
$$
whereas for $Q\ll 1$ and $\tau\gg 1$ simultaneously we get
$$
\Gamma_C(Q,\tau,\mu)=\tau_1(1)e^{-\Qabs\tau/\sqrt{\tau^*}}
\left(1+{\Qabs\over\sqrt{\tau^*}}
\big(\mu(\tau^*-1)-\tau_0\tau^*\big)+\O(Q^2)\right)
.\eqno(2.57)
$$

The universal peak of the backscattering cone is then evaluated by inserting
the estimate (2.56) into eq. (2.48), using eqs. (2.29, 31).
We thus obtain the following expression
$$
\gamma_C(Q)=\gamma(1,1)\bigg(1-{\Qabs\over\Delta Q}+\O(Q^2)\bigg)
,\eqno(2.58)
$$
where the width of the cone reads
$$
\Delta Q={3\gamma(1,1)\over\tau_1(1)^2}\,{1\over\sqrt{\tau^*}}
,\eqno(2.59)
$$
i.e., in physical units
$$
\Delta\theta={3\gamma(1,1)\over\tau_1(1)^2}\,{1\over k_1\sqrt{\ell\ell^*}}
,\eqno(2.60)
$$
with $k_1$ being the wavenumber of radiation outside the diffusive medium.
The $1/\sqrt{\ell\ell^*}$ scaling law of the width of the backscattering cone
is one of the most striking results of this paper.
We shall come back to this point in detail in section 5.

\medskip
\noindent{\bf 2.6. Extinction and absorption lengths}

Up to now we have assumed that the diffusive medium is conservative.
This means that the light only experiences elastic collisions;
there is neither absorption, nor inelastic scattering,
implying the normalization (2.2) of the phase function.
We now want to discuss briefly the case of a weakly absorbing diffusive medium,
characterized by a non-trivial albedo $a$ such that $1-a\ll 1$.
In this case the diffuse intensity is expected to die off exponentially
inside the medium, with a characteristic absorption length $L_{\rm abs}$.

The known expression [2, 3, 9] of the absorption length can easily
be recovered by means of the formalism exposed in the previous section,
in the case of general anisotropic scattering
and in the regime of weak absorption.
We shall actually determine the extinction length
of the more general problem defined by the $Q$-dependent
Schwarzschild-Milne equation (2.49).
We look for a slowly varying source function of the form (2.52),
along the lines of section 2.5.
The main difference is that we have now $\varpi_0=a\ne 1$.
We expand the recursion relations (2.53) for $\ell=0$ and $\ell=1$ as follows
$$
\left({1\over\varpi_0}-1\right)c_0
={s_0^2-Q^2\over 3}c_0+{s_0\over 3}c_1+\cdots,\quad
\left({1\over\varpi_1}-1\right)c_1=s_0c_0+\cdots
\eqno(2.61)
$$
from which we obtain
$$
s_0^2\approx {Q^2+3(1-a)\over\tau^*}
.\eqno(2.62)
$$
The $Q$-dependent extinction length reads
$$
L_{\rm ext}(Q)\approx{\ell\over s_0}\approx\left({\ell\ell^*\over Q^2+3(1-a)}
\right)^{1/2}\quad(Q\ll1,\,\,1-a\ll1)
.\eqno(2.63)
$$
The absorption length is obtained by setting $Q=0$
in the above expression, namely
$$
L_{\rm abs}\approx\left({\ell\ell^*\over3(1-a)}\right)^{1/2}\quad(1-a\ll1)
,\eqno(2.64)
$$
in agreement with refs. [2, 3, 9].

\medskip
\noindent{\bf 3. LARGE INDEX MISMATCH REGIME}

In this section, we extend to the general case of anisotropic scattering
the approach of ref. [12], which predicts the behavior of physical quantities
in the regime where the optical indices $n_0$ and $n_1$
of the diffusive medium and of the surroundings
are very different from each other,
i.e., when their ratio $m=n_0/n_1$ is either very small, or very large.
As already pointed out in ref. [12],
important simplifications occur in these regimes
of a large index mismatch,
where the mean transmission coefficient of the boundaries of the medium
is very small.
To be more specific, radiation cannot enter the medium
(respectively, leave the medium)
in the limit $m\ll 1$ (respectively, $m\gg 1$),
except at normal incidence.
Ref. [9] already contains part of the novel results of this section.

\medskip
\noindent{\bf 3.1. Diffuse reflection and transmission}

Along the lines of ref. [12],
we evaluate the reflected and transmitted intensities
in the large index mismatch regime by means of the following
singular perturbative expansion
of the Green's function $G_S(\tau,\mu,\tau',\mu')$, defined by eq. (2.26).

The starting point consists in noticing the identity
$$
M_L(\tau,\mu,\tau',\mu')=R(-\mu')M_B(\tau,\mu,-\tau',-\mu')
\eqno(3.1)
$$
between both kernels defined in eq. (2.23).
Using $R(\mu)=1-T(\mu)$, we can recast eq. (2.26) as
$$
\eqalign{
G_S(\tau,\mu,\tau',\mu')&=p_0(\mu,\mu')\delta(\tau-\tau')\cr
&+\int_0^\infty\d\tau''\int_{-1}^1{\d\mu''\over 2}
\big[M_B(\tau-\tau'',\mu,0,\mu'')+M_B(\tau+\tau'',\mu,0,-\mu'')\big]\cr
&{\hskip 3.25cm}\times G_S(\tau'',\mu'',\tau',\mu')\cr
&-\int_0^\infty\d\tau''\int_{-1}^1{\d\mu''\over 2}T(\mu'')
M_B(\tau,\mu,\tau'',-\mu'')G_S(\tau'',\mu'',\tau',\mu').\cr
}
\eqno(3.2)
$$

In the limit of an infinitely strong index mismatch,
i.e., for $m=0$ or $m=\infty$,
the transmission coefficient $T(\mu)$ vanishes identically,
so that the last integral of eq. (3.2), involving $T(\mu)$, is absent.
It can be checked that the remaining terms only determine
the Green's function up to an additive constant.
This constant is only fixed by the integral involving $T(\mu)$ in eq. (3.2).
It can therefore be expected to diverge as $m\to 0$ and $m\to\infty$.

In order to demonstrate this explicitly,
we expand the Green's function according to
$$
G_S(\tau,\mu,\tau',\mu')=C_S+G_0(\tau,\mu,\tau',\mu')+\cdots
,\eqno(3.3)
$$
with the hypothesis that $C_S$ diverges,
whereas $G_0(\tau,\mu,\tau',\mu')$ remains finite,
and the dots stand for terms which go to zero, as $m\to 0$ or $m\to\infty$.
The {\it finite part} $G_0(\tau,\mu,\tau',\mu')$ obeys the following equation
$$
\eqalign{
G_0(\tau,\mu,\tau',\mu')&=p_0(\mu,\mu')\delta(\tau-\tau')\cr
&+\int_0^\infty\d\tau''\int_{-1}^1{\d\mu''\over 2}
\big[M_B(\tau-\tau'',\mu,0,\mu'')+M_B(\tau+\tau'',\mu,0,-\mu'')\big]\cr
&{\hskip 3.25cm}\times G_0(\tau'',\mu'',\tau',\mu')\cr
&-C_S\int_0^\infty\d\tau''\int_{-1}^1{\d\mu''\over 2}T(\mu'')
M_B(\tau,\mu,\tau'',-\mu''),\cr
}
\eqno(3.4)
$$
together with the consistency condition
$$
\int_0^\infty\d\tau\int_0^\infty\d\tau'
\int_{-1}^1 {\d\mu\over 2}\int_{-1}^1 {\d\mu'\over 2}
T(\mu')M_B(\tau+\tau',\mu,0,-\mu')G_0(\tau',\mu',\tau'',\mu'')=0
,\eqno(3.5)
$$
derived along the lines of ref. [12].

The constant $C_S$ of the expansion (3.3)
can be derived by integrating eq. (3.4) over
the variables $0<\tau<\infty$ and $-1<\mu<1$.
This yields
$$
C_S={4\over\T}
,\eqno(3.6)
$$
where $\T$ is the {\it mean flux transmission coefficient}
$$
\T=2\int_0^1\mu T(\mu)\d\mu=\left\{\matrix{
\frc{4m(m+2)}{3(m+1)^2}\hfill&(m\le 1),\hfill\cr\null\cr
\frc{4(2m+1)}{3m^2(m+1)^2}\hfill&(m\ge 1).\hfill\cr}\right.
\eqno(3.7)
$$
This quantity assumes its maximum $\T=1$ in the absence of any index mismatch,
i.e., for $m=1$, and it vanishes in both cases of a large index mismatch,
according to
$$
\T\approx\left\{\matrix{
\frc{8m}{3}&(m\ll 1),\hfill\cr\null\cr
\frc{8}{3m^3}&(m\gg 1).\hfill\cr}\right.
\eqno(3.8)
$$

The asymptotic behavior in the limits $m\ll 1$ and $m\gg 1$
of the quantities pertaining to reflection and transmission
is immediately obtained by replacing in eqs. (2.29--31)
the Green's function $G_S(\tau,\mu,\tau',\mu')$
by its leading constant term $C_S$.
We thus obtain
$$
\tau_0\approx{4\over 3\T},
\quad
\tau_1(\mu)\approx{4\mu\over\T},
\quad
\gamma(\mu_a,\mu_b)\approx{4\mu_a\mu_b\over\T}
.\eqno(3.9)
$$
These predictions are identical to those derived
in ref. [12], in the case of isotropic scattering.
We have thus shown that the quantities
which determine the diffuse reflected and transmitted intensities
do not depend {\it at all} on the anisotropy of the cross-section
in the large index mismatch limit.

\medskip
\noindent{\bf 3.2. Enhanced backscattering cone}

The shape $\gamma_C(Q)$ of the cone of enhanced backscattering
for a normal incidence can also be evaluated analytically in the regimes
$m\ll 1$ or $m\gg 1$ of a large index mismatch.
To do so, we recast the $Q$-dependent Schwarzschild-Milne equation (2.49)
in a form similar to eq. (3.2), namely
$$
\eqalign{
\Gamma_C(Q,\tau,\mu)&=p_0(\mu,1)e^{-\tau}\cr
&+\int_0^\infty\d\tau'\int_{-1}^1{\d\mu'\over 2}
\big[M_B(Q,\tau-\tau',\mu,0,\mu')+M_B(Q,\tau+\tau',\mu,0,-\mu')\big]\cr
&{\hskip 3.25cm}\times\Gamma_C(Q,\tau',\mu')\cr
&-\int_0^\infty\d\tau'\int_{-1}^1{\d\mu'\over 2}T(\mu')
M_B(Q,\tau,\mu,\tau',-\mu')\Gamma_C(Q,\tau',\mu').\cr
}
\eqno(3.10)
$$

By inserting the estimates (3.9) into the general result (2.59),
we anticipate that the width of the cone is small in the large index mismatch
regime, since it is proportional to the mean transmission $\T$.
Our present goal is to determine the scaling law
of the backscattering cone $\gamma_C(Q)$
in the scaling regime where both $Q$ and $\T$ are simultaneously small.
Under these circumstances, we look for a solution to eq. (3.10)
of the form
$$
\Gamma_C(Q,\tau,\mu)\approx\gamma_C(Q)e^{-\vert Q\vert\tau/\sqrt{\tau^*}}
,\eqno(3.11)
$$
where the extinction length of the exponential decay
has been taken from eq. (2.57).
We insert this form into eq. (3.10), and we integrate again both sides over
the variables $0<\tau<\infty$ and $-1<\mu<1$.
The integrals independent of the transmission $T(\mu)$
can be performed explicitly, by means of the formula
$$
\int_{-1}^1{\d\mu\over 2}\big(1+Q^2(1-\mu^2)\big)^{-1/2}
={\arctan Q\over Q}\approx 1-Q^2/3\quad(Q\ll 1)
,\eqno(3.12)
$$
whereas the $Q$-dependence of the integral involving $T(\mu)$
can be neglected in the scaling regime.
After some algebra we are left with the following result
$$
\gamma_C(Q)\approx{1\over\vert Q\vert\sqrt{\tau^*}/3+\T/4}
\quad(Q\ll 1,\T\ll 1)
.\eqno(3.13)
$$
The small-$Q$ expansion of this prediction reads
$$
\gamma_C(Q)
\approx{4\over\T}\bigg(1-{4\vert Q\vert\sqrt{\tau^*}\over 3\T}+\cdots\bigg)
.\eqno(3.14)
$$
The width of the cone therefore scales as
$$
\Delta Q\approx{3\T\over 4\sqrt{\tau^*}}
,\eqno(3.15)
$$
in agreement with the general formula (2.59), together with the results (3.9).

\medskip
\noindent{\bf 4. VERY ANISOTROPIC SCATTERING}

In this section we investigate the regime where the scattering cross-section
is very anisotropic, i.e., strongly peaked in a narrow cone
of width $\Thetarms\ll 1$ around the forward direction,
with $\Thetarms^2=\langle\Theta^2\rangle$.
We thus have $1-\varpi_1\approx\Thetarms^2/2\ll 1$,
so that the anisotropy parameter $\tau^*\approx 2/\Thetarms^2$ is very large.
The wavelength $\lambda_0$ of radiation in the medium,
the scattering mean free path $\ell$,
and the transport mean free path $\ell^*$
can therefore be considered as three
independent length scales $(\lambda_0\ll\ell\ll\ell^*)$,
besides other characteristic lengths,
such as the sample thickness $L$,
and possibly the absorption length $L_{\rm abs}$.

The interest in this very anisotropic regime is twofold.
First, we shall show that the Schwarzschild-Milne problem
in the absence of internal reflections is exactly solvable in this regime,
just as it is for isotropic scattering,
whereas the intermediate situation of a general anisotropy
can only be treated numerically.
Second, the dependence of physical quantities on anisotropy,
with respect to the well-known isotropic case,
can be expected to yield the largest effects in the very anisotropic regime.
We shall come back to this point in section 5.

\medskip
\noindent{\bf 4.1. The example of large spheres}

We first show that very anisotropic scattering can be realized experimentally.
We consider the multiple scattering of light by large dielectric spheres,
with radius $a$ much larger than the wavelength $\lambda_0$ in the medium,
i.e., with scale parameter $k_0a\gg 1$, so that geometrical optics can be used.
Furthermore we assume that the optical index $n_S$ of the spheres
is very close to the mean index $n_0$ of the medium, i.e.,
$$
{n_S\over n_0}=m_S=1+\delta_S\quad(\vert\delta_S\vert\ll 1)
.\eqno(4.1)
$$
It is expected that scattering mostly takes place
in the forward direction in this regime.

The study of the scattering cross-section of spheres
is an old classical subject.
The full solution was first derived by Mie in 1908.
Ref. [25] provides an extensive overview of this field.
The regime $k_0a\gg 1$ and $\vert\delta_S\vert\ll 1$
still has to be split into several subcases,
according to the value of the combination $\vert\delta_S\vert k_0a$.
This can be understood as follows.
In the framework of geometrical optics
we can distinguish between the diffracted light,
which is outgoing within an angle $\Theta_{\rm diffr}\sim 1/(k_0a)$,
independent of $\delta_S$, and the refracted light,
which is outgoing within an angle $\Theta_{\rm refr}\sim\vert\delta_S\vert$,
independent of $k_0a$.

 From now on we concentrate our attention on the regime
$\vert\delta_S\vert\ll 1$, $k_0a\gg 1$, and $\vert\delta_S\vert k_0a\gg 1$.
In this regime we have $\Theta_{\rm diffr}\ll\Theta_{\rm refr}\ll 1$.
The cross-sections associated with each of the above processes asymptotically
reads $\sigma_{\rm diffr}\approx\sigma_{\rm refr}\approx\pi a^2$,
so that the total elastic cross-section $\sigma\approx 2\pi a^2$
is twice the geometrical one.
This is the celebrated {\it extinction paradox}.
We make the following approximations.
We neglect the diffraction phenomenon by setting $\Theta_{\rm diffr}=0$.
We treat the refracted light according to the laws of geometrical optics,
as illustrated in Figure 1.
We neglect all the rays which are reflected at least once
at the surface of the sphere, so that we only have to consider
the refracted ray drawn on the Figure.
The corresponding phase function reads
$$
p_{\rm refr}(\Theta)={4\sin\beta\cos\beta\over\sin\Theta}
\left\vert{\d\beta\over\d\Theta}\right\vert
,\eqno(4.2)
$$
with the notations of Figure 1, which also imply
$$
\Theta\approx -2\delta_S\tan\beta
.\eqno(4.3)
$$
Eqs. (4.2, 3) lead to the Lorentzian-squared scaling form [25]
$$
p_{\rm refr}(\Theta)\approx{16\delta_S^2\over(\Theta^2+4\delta_S^2)^2}
.\eqno(4.4)
$$
The cross-section is thus strongly peaked in the forward direction,
as anticipated.

We now turn to the evaluation of
the coefficient $\varpi_{1,\rm refr}$ corresponding to refracted light.
The integral $\langle\Theta^2\rangle_{\rm refr}
=\int_{-\infty}^\infty\Theta^2 p_{\rm refr}(\Theta)\,\Theta\d\Theta$,
with the phase function of eq. (4.4), is logarithmically divergent.
A more careful treatment is therefore needed,
which consists in using directly eq. (4.2),
choosing $u=\sin^2\beta$ as integration variable.
We thus obtain
$$
\varpi_{1,\rm refr}={1\over 3m_S^2}+{4\over m_S^2}\int_0^{u_{\rm max}}
u\d u\sqrt{(1-u)(m_S^2-u)}
,\eqno(4.5{\rm a})
$$
with $u_{\rm max}$ being the smaller of both numbers $1$ and $m_S^2$.
The integral in eq. (4.5a) can be performed explicitly
for both $m_S<1$ and $m_S>1$.
In both cases we obtain
$$
\varpi_{1,\rm refr}\approx 1-2\delta_S^2\ln{\Lambda\over\vert\delta_S\vert},
\quad\hbox{with}\,\Lambda=2e^{-3/2}\quad(\vert\delta_S\vert\ll 1)
.\eqno(4.5{\rm b})
$$

The anisotropy parameter can be derived from the above estimate,
not forgetting that $\sigma_{\rm diffr}\approx\sigma_{\rm refr}
\approx\sigma/2$ implies $1-\varpi_1=(1-\varpi_{1,\rm refr})/2$.
To leading order for $\vert\delta_S\vert\ll 1$, this yields
$$
\tau^*={\ell^*\over\ell}\approx{1\over\delta_S^2
\ln\frc{\Lambda}{\vert\delta_S\vert}}
.\eqno(4.6)
$$
It is therefore evident that large spheres with a weak dielectric contrast
provide a physical instance of very anisotropic scattering,
as described in the beginning of this section,
with the refraction angle $\Theta_{\rm refr}\sim\vert\delta_S\vert\ll 1$
playing the role of $\Thetarms$, up to a logarithmic correction.
We shall come back to this last point in section 5.

\medskip
\noindent{\bf 4.2. Scaling limit of the Schwarzschild-Milne equation}

We now show that the general formalism of section 2
undergoes important simplifications in the regime of very
anisotropic scattering.
These will allow us to derive analytical results for
physical quantities, including the shape of the backscattering cone,
in the absence of index mismatch at the interface.
This solution, to be described in sections 4.3 and 4.4,
therefore shows for the first time that both limiting cases
of isotropic scattering and of very anisotropic scattering
are nearly on the same footing as far as the existence
of exact results is concerned.

The simplifications in the regime of very anisotropic scattering
can be understood in physical terms as follows.
Consider the random sequence of scattering events experienced by a light ray.
At every scattering event, the direction $\n$ of the ray is only modified
by a slight amount of order $\Thetarms$, with $\Thetarms^2\approx 2/\tau^*$.
The extremity of the vector $\n$ therefore performs a Brownian motion
on the unit sphere, with a small angular diffusivity
$\varepsilon\sim\Thetarms^2$.
A similar picture has been used for long in the context
of semiflexible polymer chains:
the so-called Kratky-Porod theory models polymers as continuous persistent
walks, whose unit tangent vectors obey a diffusion equation on the sphere
(see ref. [26] for a review).

In more quantitative terms, the relationship (2.7) between the $\n$-dependence
of the specific intensity $I(\r,\n)$ and the source function $\Gamma(\r,\n)$
takes the form
$$
\Gamma(\r,\n)\approx(1+\varepsilon\Delta)I(\r,\n)
,\eqno(4.7)
$$
where $\Delta$ is the Legendre operator (Laplace operator on the unit sphere)
$$
\Delta=\cot\theta{\p\over\p\theta}+{\p^2\over\p\theta^2}
+{1\over\sin^2\theta}{\p^2\over\p\varphi^2}
.\eqno(4.8)
$$
As a consequence, the linear operator $\D$ introduced in eq. (2.11) reads
$$
\D\approx 1+\varepsilon\Delta_0
,\eqno(4.9)
$$
where
$$
\Delta_0=\cot\theta{\p\over\p\theta}+{\p^2\over\p\theta^2}
=(1-\mu^2){\p^2\over\p\mu^2}-2\mu{\p\over\p\mu}
\eqno(4.10)
$$
is the Legendre operator (4.8) in the $\varphi$-independent sector.

The angular diffusivity $\varepsilon$
is then determined by observing that $P_1(\mu)=\mu$
is an eigenfunction of the operator $\Delta_0$, with eigenvalue $(-2)$,
and of the operator $\D$, with eigenvalue $\varpi_1$, by virtue of eq. (2.14).
We thus identify $\varpi_1=1-2\varepsilon$, whence, using eq. (2.16),
$$
\varepsilon={1\over 2\tau^*}={\ell\over 2\ell^*}\approx{\Thetarms^2\over 4}
,\eqno(4.11)
$$
in agreement with the above heuristic estimate.

The radiative transfer equation (2.8) thus assumes the differential form
$$
2\tau^*\mu{\d\over\d\tau} I(\tau,\mu)=\Delta_0 I(\tau,\mu)
,\eqno(4.12)
$$
which demonstrates that physical quantities vary
over length scales of order $\tau\sim\tau^*$, i.e., $z\sim\ell^*$.
In the next two sections we present our exact analytical predictions
concerning physical observables pertaining to optically thick slabs,
in the regime of very anisotropic scattering,
in the absence of internal reflections.
The analysis roughly follows the presentation of section III of ref. [12],
devoted to the case of isotropic scattering.

\medskip
\noindent{\bf 4.3. Exact treatment in the absence of internal reflections:
homogeneous Schwarzschild-Milne equation}

This section is devoted to an analytic determination
of the solution $\Gamma_H(\tau,\mu)$ of the homogeneous
Schwarzschild-Milne equation in the regime of very anisotropic scattering,
and of the associated quantities of interest, $\tau_0$ and $\tau_1(\mu)$.

It is convenient to consider the Laplace transform $g_H(s,\mu)$
of $\Gamma_H(\tau,\mu)$, defined as follows
$$
g_H(s,\mu)=\int_0^\infty\d\tau\Gamma_H(\tau,-\mu)e^{s\tau}\quad (\Re s<0)
.\eqno(4.13)
$$
The Schwarzschild-Milne equation (2.24) can be recast as
$$
g_H(s,\mu)=\left\{\matrix{
\D\left[\frc{g_H(s,\mu)}{1+s\mu}\right]\hfill&(-1<\mu<0),\hfill\cr\null\cr
\D\left[\frc{g_H(s,\mu)-\tau^*\tau_1(\mu)/3}{1+s\mu}\right]\hfill
&(0<\mu<1),\hfill\cr}\right.
\eqno(4.14)
$$
since we have $g_H(-1/\mu,\mu)=\tau^*\tau_1(\mu)/3$ for $0<\mu<1$,
as a consequence of eqs. (2.31, 35).

We now expand eq. (4.14), using the expression (4.9, 11)
of the operator $\D$.
We introduce the rescaled Laplace variable
$$
\sigma=2s\tau^*
,\eqno(4.15)
$$
as well as a new unknown function $h_H(\sigma,\mu)$, defined as follows
$$
g_H(s,\mu)=\left\{\matrix{
4\tau^{*2}(1+s\mu)h_H(\sigma,\mu)\hfill&(-1<\mu<0),\hfill\cr
4\tau^{*2}(1+s\mu)h_H(\sigma,\mu)+\tau^*\tau_1(\mu)/3\hfill&(0<\mu<1).
\hfill\cr
}\right.
\eqno(4.16)
$$
We assume that $\tau_1(\mu)$ vanishes faster than linearly in $\mu$
for $\mu\to 0$.
This hypothesis will be checked a posteriori,
as it will turn out that $\tau_1(\mu)\sim\mu^{3/2}$ for $\mu\to 0$
(see eq. (4.53)).
The function $h_H(\sigma,\mu)$ is then
continuously differentiable as a function of $\mu$,
and it obeys the following equation
$$
(\Delta_0-\sigma\mu)h_H(\sigma,\mu)=\left\{\matrix{
0\hfill&(-1<\mu<0),\hfill\cr
\tau_1(\mu)/6\hfill&(0<\mu<1).\hfill\cr}\right.
\eqno(4.17)
$$

By means of the change of variable
$$
\mu=\tanh x
,\eqno(4.18)
$$
eq. (4.17) can be recast in the form of the following inhomogeneous
Schr\"odinger equation on the real $x$-line
$$
h_H''(\sigma,x)-\sigma V(x)h_H(\sigma,x)=\left\{\matrix{
0\hfill&(x< 0),\hfill\cr
V(x)\nu(x)/6\hfill&(x\ge 0).\hfill\cr}\right.
\eqno(4.19)
$$
In this formula the accents denote differentiations with respect to $x$.
The potential
$$
V(x)=\tanh x(1-\tanh^2x)
\eqno(4.20)
$$
is an odd function of $x$, such that $V(x)\d x=\mu\d\mu$, and we have set
$$
\tau_1(\mu)=\mu\nu(x)\quad(0<\mu<1,\quad x>0)
.\eqno(4.21)
$$

Equation (4.19) is a self-consistent equation for the two unknown functions
$h_H(\sigma,x)$ and $\nu(x)$.
Analyticity properties in the complex $\sigma$-variable turn out to
allow for an exact analytical solution of this equation.

We introduce a basis of two elementary solutions
$\{u_1(\sigma,x),u_2(\sigma,x)\}$
of the (homogeneous) Schr\"odinger equation
$$
u''(\sigma,x)-\sigma V(x)u(\sigma,x)=0
,\eqno(4.22)
$$
with the following asymptotic behavior
$$
u_1(\sigma,x)\approx 1,\quad u_2(\sigma,x)\approx x\quad (x\to -\infty)
,\eqno(4.23)
$$
up to exponentially small corrections.
Similarly we introduce the basis $\{v_1(\sigma,x),v_2(\sigma,x)\}$,
with the asymptotic behavior
$$
v_1(\sigma,x)\approx 1,\quad v_2(\sigma,x)\approx x\quad (x\to\infty)
.\eqno(4.24)
$$
The Schr\"odinger equation (4.22) admits solutions
with the above boundary conditions,
since the potential $V(x)$ vanishes exponentially as $x\to\pm\infty$.
The four above solutions are entire functions of $\sigma$,
i.e., they are analytic in the whole $\sigma$-plane.
Furthermore they are related by the following identities
$$
v_1(\sigma,x)=u_1(-\sigma,-x),\quad v_2(\sigma,x)=-u_2(-\sigma,-x)
,\eqno(4.25)
$$
because the potential $V(x)$ is odd.

We recall that, if $u(x)$ and $v(x)$ are any two solutions
of the Schr\"odinger equation (4.22), with the same $\sigma$,
their {\it Wronskian}
$$
W\{u,v\}=u(x)v'(x)-u'(x)v(x)
\eqno(4.26)
$$
is independent of $x$.
The boundary conditions (4.23, 24) imply that both bases of functions
have unit Wronskian, namely
$$
W\{u_1,u_2\}=W\{v_1,v_2\}=1
.\eqno(4.27)
$$

For generic values of the parameter $\sigma$,
both bases of solutions are related by a $2\times 2$
{\it transfer matrix} of the form
$$
\pmatrix{v_1\cr v_2}=\pmatrix{F(\sigma)&G(\sigma)\cr H(\sigma)&F(-\sigma)}
\pmatrix{u_1\cr u_2}
.\eqno(4.28)
$$
The three functions which enter eq. (4.28) are entire functions of $\sigma$;
the determinant of the matrix is $F(\sigma)F(-\sigma)-G(\sigma)H(\sigma)=1$;
as a consequence of eq. (4.25),
$G(\sigma)$ and $H(\sigma)$ are even functions of $\sigma$.

The above functions also govern the non-trivial asymptotic behavior
of the bases of solutions
$$
\eqalign{
u_1(\sigma,x)&\approx F(-\sigma)-G(\sigma)x,
\quad u_2(\sigma,x)\approx -H(\sigma)+F(\sigma)x\quad(x\to\infty),\cr
v_1(\sigma,x)&\approx F(\sigma)+G(\sigma)x,
\quad v_2(\sigma,x)\approx H(\sigma)+F(-\sigma)x\quad(x\to -\infty),
}
\eqno(4.29)
$$
as well as their mixed Wronskians
$$
W\{u_1,v_1\}=G(\sigma),\,\,W\{u_1,v_2\}=F(-\sigma),
\,\,W\{u_2,v_1\}=-F(\sigma),\,\,W\{u_2,v_2\}=-H(\sigma)
.\eqno(4.30)
$$

The first terms of the Taylor expansion
of $u_1(\sigma,x)$ and $u_2(\sigma,x)$ around $\sigma=0$ read
$$
\eqalign{
&u_1(\sigma,x)=1-{\sigma\over 2}(1+\tanh x)+\cdots,\cr
&u_2(\sigma,x)=x+\sigma\left({x\over 2}(1-\tanh x)+\ln(2\cosh x)\right)+\cdots
}
\eqno(4.31)
$$
We have also determined the terms of order $\sigma^2$,
which are too lengthy expressions to be reported here.
They imply
$$
F(\sigma)=1+\sigma+\sigma^2/2+\cdots,\quad G(\sigma)=\sigma^2/3+\cdots,\quad
H(\sigma)=(7/6-\pi^2/36)\sigma^2+\cdots
\eqno(4.32)
$$
On the other hand, large values of the complex parameter $\sigma$
correspond to the {\it semi-classical} regime
for the Schr\"odinger equation (4.22),
where the behavior of the functions $u_1(\sigma,x)$ and $u_2(\sigma,x)$
can be derived by means of a W.K.B.-like approximation.
This regime is analyzed in detail in Appendix B.
Let us mention that the wavefunctions display oscillations
when either $\sigma>0$ and $x<0$ or $\sigma<0$ and $x>0$,
whereas they are growing or decaying exponentially in the other two cases.

The function $G(\sigma)$ deserves some more attention,
since it will play a central role in the following.
$G(\sigma)$ can be viewed as the {\it functional determinant}
of the Schr\"odinger equation (4.22), in the following sense.
Assume $\sigma$ is such that $G(\sigma)=0$.
We have then
$$
v_1(\sigma,x)=F(\sigma)u_1(\sigma,x),\quad
u_1(\sigma,x)=F(-\sigma)v_1(\sigma,x),\quad
F(\sigma)F(-\sigma)=1
.\eqno(4.33)
$$
In other words, for such a $\sigma$, the Schr\"odinger equation (4.22)
has a bounded solution over the whole real line.
Throughout the following, using slightly improper terms,
such a value of $\sigma$ is called an {\it eigenvalue}
of the Schr\"odinger equation,
and the corresponding function $v_1(\sigma,x)$ is referred to as
the associated {\it eigenfunction}.
The semi-classical analysis of Appendix B demonstrates
that there is an infinite sequence of real eigenvalues.
We label them by an integer $-\infty<n<\infty$,
so that $\sigma_n>0$ for $n\ge 1$, $\sigma_0=0$, and $\sigma_{-n}=-\sigma_n$.
The associated eigenfunctions $v_1(\sigma_n,x)$ are orthogonal
with respect to the following indefinite metric
$$
\int_{-\infty}^\infty v_1(\sigma_m,x)v_1(\sigma_n,x)V(x)\d x
=N_n\delta_{m, n}\quad(-\infty<m, n<\infty)
.\eqno(4.34)
$$
The squared norms $N_n$ obey the symmetry property
$$
{N_{-n}\over F(-\sigma_n)}=-{N_n\over F(\sigma_n)}
,\eqno(4.35)
$$
as a consequence of eq. (4.33).
The eigenfunction $v_1(0,x)=1$ associated with the zero mode $\sigma_0=0$
is peculiar, since its squared norm reads $N_0=0$.
As a consequence, the basis of eigenfunctions
$\{v_1(\sigma_n,x), n\ne 0\}$
only spans the set of bounded functions $f(x)$ on the real $x$-line
such that
$$
\int_{-\infty}^\infty f(x)V(x)\d x=0
.\eqno(4.36)
$$
For such functions, we have
$$
f(x)=\sum_{n\ne 0}c_nv_1(\sigma_n,x),\quad
c_nN_n=\int_{-\infty}^\infty v_1(\sigma_n,x)f(x)V(x)\d x
.\eqno(4.37)
$$
The determination of the contribution of the zero mode to functions $f(x)$
which do not obey the condition (4.36), such as e.g. a constant,
requires more care.
An elegant way of dealing with this problem consists in introducing
a deformation parameter $\q$, as we shall see in section 4.4.

It is advantageous to factor the entire function $G(\sigma)$ as follows
$$
G(\sigma)={\sigma^2\over 3}P(\sigma)P(-\sigma)
,\eqno(4.38)
$$
with the notation
$$
P(\sigma)=\prod_{n\ge 1}\left(1+{\sigma\over\sigma_n}\right)
.\eqno(4.39)
$$

The explicit solution of the inhomogeneous Schr\"odinger equation (4.19)
now goes as follows.
Since $h_H(\sigma,x)$ is a regular function of $x$ in the $x\to -\infty$ limit,
it is proportional to $u_1(\sigma,x)$ for $x<0$, namely
$$
h_H(\sigma,x)=a_H(\sigma)u_1(\sigma,x)
\quad(x<0)
.\eqno(4.40)
$$
For $x>0$ we solve eq. (4.19) by {\it varying the constants},
namely we look for a solution of the form
$$
h_H(\sigma,x)=b_H(\sigma,x)v_1(\sigma,x)+c_H(\sigma,x)v_2(\sigma,x)
\quad(x>0)
.\eqno(4.41)
$$
The unknown {\it constants} $b_H(\sigma,x)$ and $c_H(\sigma,x)$ obey the
requirements
$$
\eqalignno{
&b'_H(\sigma,x)v_1(\sigma,x)+c'_H(\sigma,x)v_2(\sigma,x)=0, &(4.42{\rm a})\cr
&b'_H(\sigma,x)v_1'(\sigma,x)+c'_H(\sigma,x)v_2'(\sigma,x)=V(x)\nu(x)/6,
&(4.42{\rm b})\cr
}
$$
where the accents again denote differentiations with respect to
$x$.
Eq. (4.42a) is a constraint imposed a priori,
in order to break the large redundance of the representation (4.41);
eq. (4.42b) is then a consequence of eq. (4.19).
Eq. (4.42) can be solved for $b'_H$ and $c'_H$, using eq. (4.27).
Integrating the expressions thus obtained, we get
$$
\eqalign{
&b_H(\sigma,x)=b_H(\sigma,\infty)+\int_x^\infty
v_2(\sigma,y)\nu(y)V(y)\d y/6,\cr
&c_H(\sigma,x)=-\int_x^\infty v_1(\sigma,y)\nu(y)V(y)\d y/6.\cr
}
\eqno(4.43)
$$
Indeed there cannot be a non-zero constant $c_H(\sigma,\infty)$,
because this would correspond to an unacceptable singular solution of the form
$h_H(\sigma,x)\sim x\sim\ln(1-\mu)$ for $\mu\to 1$.

Both expressions (4.40) and (4.41) have to match at $x=0$,
together with their first derivatives.
These two conditions determine $b_H(\sigma,0)$ and $c_H(\sigma,0)$,
and some algebra then leads to the identity
$$
G(\sigma)a_H(\sigma)=-c_H(\sigma,0)
=\int_0^\infty v_1(\sigma,x)\nu(x)V(x)\d x/6
.\eqno(4.44)
$$

We can argue on eq. (4.44) as follows.
Since $g_H(s,\mu)$ is analytic in the half-plane $\Re s<0$,
$a_H(\sigma)$ has the same property for $\Re\sigma<0$,
so that the zeroes $-\sigma_n$ of $G(\sigma)$ cannot be poles of $a_H(\sigma)$.
They are therefore zeroes of $c_H(\sigma,0)$.
On the other hand, the $\sigma_n$'s are not zeroes of $c_H(\sigma,0)$,
since the integral expression in eq. (4.44) is positive for $\sigma>0$.
Hence they are poles of $a_H(\sigma)$.
The final step concerns the small-$\sigma$ behavior of the quantities
involved in eq. (4.44).
The asymptotic behavior (2.25) of $\Gamma_H(\tau,\mu)$ yields the following
double-pole structure for its Laplace transform near $s=0$
$$
g_H(s,\mu)={1\over s^2}-{\tau^*(\tau_0+\mu)\over s}+\O(1)\quad(s\to 0)
,\eqno(4.45)
$$
which implies the following small-$\sigma$ behavior of $a_H(\sigma)$
$$
a_H(\sigma)={1\over\sigma^2}+{1-\tau_0\over 2\sigma}+\O(1)\quad(\sigma\to 0)
.\eqno(4.46)
$$
We thus obtain finally
$$
a_H(\sigma)={1\over\sigma^2P(-\sigma)},\quad
-c_H(\sigma,0)={P(\sigma)\over 3}
.\eqno(4.47)
$$
Eq. (4.47) can be considered as an explicit result.
Indeed $P(\sigma)$, defined in eq. (4.39),
is for all purposes a known function,
since the eigenvalues $\sigma_n$ can be determined numerically,
essentially with arbitrary accuracy, via the partial-wave expansion
procedure of Appendix A.
Furthermore the semi-classical analysis of Appendix B
determines the asymptotic behavior of the eigenvalues
in the regime of large quantum numbers $(n\gg 1)$.

As a first consequence of the above results,
we can determine the reduced thickness $\tau_0$ of a skin layer,
i.e., the reduced extrapolation length,
by comparing the small-$\sigma$ behavior of the expression (4.47)
for $a_H(\sigma)$ with the expansion (4.46).
We thus obtain
$$
\tau_0=1-2\sum_{n\ge 1}{1\over\sigma_n}=0.71821164
.\eqno(4.48)
$$
The series converges, since $\sigma_n$ grows as $n^2$,
according to the semi-classical estimate (B.21).
The number given in eq. (4.48), as well as all the subsequent ones,
has been obtained by means of the partial-wave expansion
described in Appendix A.
This number gives an idea of the accuracy of this approach.
The most significant numerical results are listed in Table 2,
together with their counterparts in the case of isotropic scattering.

Second, the determination of $\tau_1(\mu)$ goes as follows.
We recall that this quantity is related to $\nu(x)$ by eq. (4.21).
Eqs. (4.44, 47) yield
$$
\int_0^\infty v_1(\sigma,x)\nu(x)V(x)\d x=2P(\sigma)
.\eqno(4.49)
$$

The small-$\sigma$ expansion (4.31) of $u_1(\sigma,x)$,
together with eq. (4.25),
allows us to recover the sum rules (2.34, 37b) in the following form
$$
\int_0^\infty\nu(x)V(x)\d x=2,
\quad\int_0^\infty\nu(x)\tanh xV(x)\d x=2\tau_0
.\eqno(4.50)
$$

On the other hand, the semi-classical estimates (B.18, 20) of $v_1(\sigma,x)$
and $P(\sigma)$ for large values of $\sigma=K^2$ imply
$$
\int_0^\infty\nu(x)\Ai(K^{2/3}x)x\d x\approx(6/\pi)^{1/2}K^{-5/3}
\quad(K\gg 1)
.\eqno(4.51)
$$
This estimate yields, by means of eq. (B.25) for $s=3/2$,
the small-$x$ behavior of $\nu(x)$, i.e.,
$$
\nu(x)\approx 6(2x/\pi)^{1/2}\quad(x\ll 1)
.\eqno(4.52)
$$

We thus obtain the following universal scaling behavior
of the $\tau_1$-function in the case of very anisotropic scattering
$$
\tau_1(\mu)\approx 6(2/\pi)^{1/2}\mu^{3/2}\quad(\mu\ll 1)
.\eqno(4.53)
$$
This novel result is in contrast with the linear behavior
$\tau_1(\mu)\approx\mu\sqrt{3}$ observed for isotropic scattering.
The law (4.53) confirms the hypothesis made
in the beginning of this section,
namely that $\tau_1(\mu)$ vanishes faster than linearly in $\mu$.

Finally, we can extract the full functions $\nu(x)$ and $\tau_1(\mu)$
from eq. (4.49) by means of the inversion formula (4.37).
We obtain
$$
\nu(x)=3(\tau_0+\tanh x)+2\sum_{n\ge 1}{P(\sigma_n)\over N_n}v_1(\sigma_n,x)
,\eqno(4.54)
$$
i.e.,
$$
{\tau_1(\mu)\over\mu}=
3(\tau_0+\mu)+2\sum_{n\ge 1}{P(\sigma_n)\over N_n}v_1(\sigma_n,\arg\tanh\mu)
.\eqno(4.55)
$$

The semi-classical analysis of Appendix B
implies that the contribution of the $n$-th mode to the results (4.54, 55)
falls off exponentially with $n$ for $x>0$,
according to $\exp\big[-K_n\big(I-I(x)\big)\big]$.
This exponential convergence disappears as $x\to 0$,
where eqs. (4.52, 53) apply.
The explicit terms in front of the sums in eqs. (4.54, 55),
corresponding to the contribution of the zero mode $(n=0)$,
have been anticipated from results to be derived in section 4.4.

Figure 2 shows a plot of the function $\tau_1(\mu)$,
for both isotropic scattering [12]
and very anisotropic scattering (eq. (4.55)).
The maximal values $\tau_1(1)$ for both cases are given in Table 2.
The difference between both limiting cases is remarkably small.

\medskip
\noindent{\bf 4.4. Exact treatment in the absence of internal reflections:
inhomogeneous Schwarzschild-Milne equation and backscattering cone}

In this section we derive analytical expressions
in the regime of very anisotropic scattering
of the special solution $\Gamma_S(\tau,\mu,\mu_a)$,
with its by-product the bistatic coefficient $\gamma(\mu_a,\mu_b)$,
and of the source function $\Gamma_C(Q,\tau,\mu)$,
yielding the shape $\gamma_C(Q)$ of the enhanced backscattering cone
at normal incidence.

It is advantageous to view the above quantities
as two different particular cases of the following more general problem,
which has no direct physical interpretation.
We consider the $Q$-dependent source function $\Gamma_C(Q,\tau,\mu,\mu_a)$,
which obeys the inhomogeneous $Q$-dependent Schwarzschild-Milne equation
(2.49), with the source term of eq. (2.22),
and we define the associated $Q$-dependent bistatic coefficient as
$$
\gamma_C(Q,\mu_a,\mu_b)
=\int_0^\infty\d\tau e^{-\tau/\mu_b}\Gamma_C(Q,\tau,-\mu_b,\mu_a)
.\eqno(4.56)
$$
Notations are consistent, in the sense that we have
$\gamma(\mu_a,\mu_b)=\gamma_C(Q=0,\mu_a,\mu_b)$
and $\gamma_C(Q)=\gamma_C(Q,\mu_a=\mu_b=1)$.
Among other advantages, this approach will treat in an elegant way
the problem of the zero mode with vanishing norm of the
Schr\"odinger equation (4.22).

In analogy with eq. (4.13), we define the Laplace transform
of the $Q$-dependent source function as follows
$$
g_C(Q,s,\mu,\mu_a)=\int_0^\infty\d\tau\Gamma_C(Q,\tau,-\mu,\mu_a)e^{s\tau}
\quad(\Re s<0)
.\eqno(4.57)
$$
The Schwarzschild-Milne equation (2.49) can be recast as
$$
g_C(Q,s,\mu,\mu_a)=\left\{\matrix{
\frc{\mu_ap_0(-\mu,\mu_a)}{w(Q,s,-\mu_a)}
+\D\bigg[\frc{g_C(Q,s,\mu,\mu_a)}{w(Q,s,\mu)}\bigg]
\hfill&(-1<\mu<0),\hfill\cr\null\cr
\D\bigg[\frc{g_C(Q,s,\mu,\mu_a)-\gamma_C(Q,\mu,\mu_a)}{w(Q,s,\mu)}\bigg]
\hfill&(0<\mu<1),\hfill\cr
}\right.
\eqno(4.58)
$$
with the definition
$$
w(Q,s,\mu)=\sqrt{(1+s\mu)^2+Q^2(1-\mu^2)}
.\eqno(4.59)
$$

In analogy with section 4.3, we expand eq. (4.58), using eqs. (4.9, 11).
We introduce the rescaled variables
$$
\sigma=2s\tau^*,\quad\q=Q\sqrt{\tau^*}=k_1\sqrt{\ell\ell^*}\theta
,\eqno(4.60)
$$
assuming $\q\ge0$,
as well as a new unknown function $h_C(\q,\sigma,\mu,\mu_a)$,
defined as follows
$$
g_C(Q,s,\mu,\mu_a)=\left\{\matrix{
2\tau^*w(Q,s,\mu)h_C(\q,\sigma,\mu,\mu_a)\hfill&(-1<\mu<0),\hfill\cr
2\tau^*w(Q,s,\mu)h_C(\q,\sigma,\mu,\mu_a)+\gamma_C(Q,\mu,\mu_a)
\hfill&(0<\mu<1).\hfill\cr
}\right.
\eqno(4.61)
$$

The definition (4.60) of the rescaled variable $\q$ is in full accord with
the general prediction (2.59) for the width of the backscattering cone.
In more precise terms, and in analogy with eq. (2.58), we have
$$
\gamma_C(Q,\mu_a,\mu_b)=\gamma(\mu_a,\mu_b)
-{\q\over 3}\tau_1(\mu_a)\tau_1(\mu_b)+\O(\q^2)\quad(\q\ll 1)
.\eqno(4.62)
$$

We assume that $\gamma_C(Q,\mu_a,\mu_b)$ vanishes faster than linearly
as $\mu_a$ or $\mu_b\to 0$.
Using the change of variable (4.18),
we are again left with an inhomogeneous Schr\"odinger equation, namely
$$
h_C''(\q,\sigma,x,x_a)-\big(\sigma V(x)+\q^2W(x)\big)h_C(\q,\sigma,x,x_a)
=\left\{\matrix{-2\mu_a\delta(x+x_a)\hfill&(x<0),\hfill\cr
\mu_a V(x)\rho(\q,x,x_a)\hfill&(x>0).\hfill\cr}\right.
\eqno(4.63)
$$
The potential has, besides the odd component $V(x)$ of eq. (4.20),
an even component
$$
W(x)=(1-\tanh^2 x)^2
,\eqno(4.64)
$$
and we have set
$$
\gamma_C(Q,\mu,\mu_a)=\mu\mu_a\rho(\q,x,x_a)
\quad(\mu=\tanh x>0,\,\,\mu_a=\tanh x_a>0)
.\eqno(4.65)
$$

The main difference introduced by the deformation parameter $\q$
resides in the spectrum of the Schr\"odinger equation
$$
u''(\sigma,x)-\big(\sigma V(x)+\q^2 W(x)\big)u(\sigma,x)=0
.\eqno(4.66)
$$
The eigenvalues with label $n\ne 0$ acquire a regular $\q$-dependence
of the form
$$
\sigma_n(\q)=-\sigma_{-n}(\q)=\sigma_n+\O(\q^2)
,\eqno(4.67)
$$
as well as the associated eigenfunctions $v_1(\q,\sigma_n(\q),x)$,
whereas the double degeneracy of the zero mode $\sigma_0=0$
is lifted into the following two exact eigenvalues and eigenfunctions
of eq. (4.66)
$$
\sigma_{\pm 0}(\q)=\pm 2\q,\quad v_1(\q,\sigma_{\pm 0}(\q),x)
=e^{\pm\q(1-\tanh x)}=e^{\pm\q(1-\mu)}
,\eqno(4.68)
$$
with squared norms
$$
N_{\pm 0}(\q)=\pm{e^{\pm 2\q}\over\q}\left({\sinh 2\q\over 2\q}-\cosh
2\q\right)
.\eqno(4.69)
$$
The introduction of labels $\pm 0$ is consistent with setting $n=0$
in formulas such as eq. (4.35) or (4.67),
rather than with the standard arithmetics of integers!

For any $\q\ne 0$, the set of eigenfunctions $\{v_1(\q,\sigma_n(\q),x)\}$,
where the label $n$ runs over the non-zero algebraic integers $(n\ne 0)$
plus both values $n=\pm 0$,
now spans the whole space of bounded functions $f(x)$ on the real line.
The difficulty of the constraint (4.36),
due to the vanishing norm of the zero mode at $\q=0$,
is thus cured in a natural way.

The mixed Wronskian $G(\q,\sigma)=W\{u_1,v_1\}$ can still be factorized
over its zeroes, in analogy with eq. (4.38), namely
$$
G(\q,\sigma)=G(\q,0)R(\q,\sigma)R(\q,-\sigma)
,\eqno(4.70)
$$
with
$$
R(\q,\sigma)=\prod_{n\ge 0}\left(1+{\sigma\over\sigma_n(\q)}\right)
.\eqno(4.71)
$$

The prefactor $G(\q,0)$ of eq. (4.70) is a non-trivial function of $\q$.
Indeed this quantity can be viewed as the functional determinant
of the Schr\"odinger equation (4.66) with $\sigma=0$, namely
$$
u''(\sigma,x)-\q^2 W(x)u(\sigma,x)=0
.\eqno(4.72)
$$
Eq. (4.72) is equivalent to the spheroidal equation
studied by Meixner and Sch\"afke [27].
Some properties of this equation have been studied in detail [28],
in an investigation of nematic phases of semiflexible polymer chains.
The occurrence of eq. (4.72) in that context is related to the Kratky-Porod
description of persistent chains, mentioned in section 4.2.

Eq. (4.72) has a discrete spectrum of imaginary
eigenvalues of the form $\q=\pm i\xi_n$ $(n\ge 1)$,
as shown by the semi-classical analysis of Appendix B.
On the other hand, the regularity of $G(\q,\sigma)$ at $\q=0$ implies
the small-$\q$ behavior $G(\q,0)\approx 4\q^2/3$, whence
$$
G(\q,0)={4\q^2\over 3}\prod_{n\ge 1}\left(1+{\q^2\over\xi_n^2}\right)
.\eqno(4.73)
$$

The solution of the inhomogeneous Schr\"odinger equation (4.63)
follows the lines of section 4.3.
The particular values $x=-x_a<0$ and $x=0$ define three sectors,
in which we look for a solution of the following form
$$
h_C(\q,\sigma,x,x_a)=\left\{\matrix{
a_C(\q,\sigma,x_a)u_1(\q,\sigma,x)\hfill&(x<-x_a),\hfill\cr
b_C(\q,\sigma,x_a)u_1(\q,\sigma,x)+c_C(\q,\sigma,x_a)u_2(\q,\sigma,x)
\hfill&(-x_a<x<0),\hfill\cr
d_C(\q,\sigma,x,x_a)v_1(\q,\sigma,x)+e_C(\q,\sigma,x,x_a)v_2(\q,\sigma,x)
\hfill&(x>0).\hfill\cr}\right. 
\eqno(4.74)
$$
The {\it constants} which enter the last of these expressions
obey the conditions
$$
\eqalign{
d_C'(\q,\sigma,x,x_a) v_1(\q,\sigma,x)+e'_C(\q,\sigma,x,x_a)
v_2(\q,\sigma,x)&=0,\cr
d_C'(\q,\sigma,x,x_a)v_1'(\q,\sigma,x)+e_C'(\q,\sigma,x,x_a)
v_2'(\q,\sigma,x)&=\mu_a V(x)\rho(\q,x,x_a),\cr
}
\eqno(4.75)
$$
whence
$$
\eqalign{
d_C(\q,\sigma,x,x_a)&=d_C(\q,\sigma,\infty,x_a)
+\mu_a\int_x^\infty v_2(\q,\sigma,y)\rho(\q,y,x_a)V(y)\d y,\cr
e_C(\q,\sigma,x,x_a)&=-\mu_a\int_x^\infty v_1(\q,\sigma,y)\rho(\q,y,x_a)
V(y)\d y.
}
\eqno(4.76)
$$
On the other hand, the matching of the solution (4.74) at $x=-x_a$ yields
$$
\eqalign{
b_C(\q,\sigma,x_a)&=a_C(\q,\sigma,x_a)+2\mu_au_2(\q,\sigma,x_a),\cr
c_C(\q,\sigma,x_a)&=-2\mu_au_1(\q,\sigma,x_a),\cr
}
\eqno(4.77)
$$
whereas its matching at $x=0$ leads to
$$
\eqalign{
d_C(\q,\sigma,0,x_a)
&=F(\q,-\sigma)b_C(\q,\sigma,x_a)-H(\q,\sigma)c_C(\q,\sigma,x_a),\cr
-e_C(\q,\sigma,0,x_a)
&=G(\q,\sigma)b_C(\q,\sigma,x_a)-F(\q,\sigma)c_C(\q,\sigma,x_a),\cr
}
\eqno(4.78)
$$
with the notations (4.28).
We are thus left with
$$
G(\q,\sigma)a_C(\q,\sigma,x_a)=-2\mu_av_1(\q,\sigma,-x_a)-e_C(\q,\sigma,0,x_a)
.\eqno(4.79)
$$
We can now follow the approach used on eq. (4.44).
Since $a_C(\q,\sigma,x_a)$ is holomorphic in the half-plane $\Re\sigma<0$,
the zeroes $-\sigma_n(\q)$ for $n\ge 0$ of $G(\q,\sigma)$
cannot be poles of $a_C(\q,\sigma,x_a)$.
Hence they are zeroes of the right-hand side of eq. (4.79).

We are therefore left with the problem of finding $e_C(\q,\sigma,0,x_a)$,
an entire function of $\sigma$, from the knowledge of its values
on the sequence of points $\sigma=-\sigma_n(\q)$ $(n\ge 0)$,
together with a natural assumption of minimal growth at infinity
compatible with these data.
This is a generalization to an entire function
of the problem of finding a polynomial $Q(z)$ with minimal degree,
knowing its values at $N$ points, namely $Q(z_n)=Q_n$ for $1\le n\le N$.
It is useful to view the $z_n$'s as the zeroes of the normalized polynomial
$$
P(z)=\prod_{1\le n\le N}(z-z_n)
.\eqno(4.80)
$$
The solution $Q(z)$ with minimal degree (generically $N-1$)
is given by the following {\it Lagrange interpolation formula}
$$
Q(z)=\sum_{1\le n\le N} Q_n\prod_{1\le m\ne n\le N}{z-z_m\over z_n-z_m}
=P(z)\sum_{1\le n\le N}{Q_n\over(z-z_n)P'(z_n)}
.\eqno(4.81)
$$

Extending eq. (4.81) to the present case of an infinite sequence of data
for an unknown entire function, we obtain
$$
-e_C(\q,\sigma,0,x_a)
=2\mu_a R(\q,\sigma)\sum_{n\ge 0}
{v_1(\q,-\sigma_n(\q),-x_a)\over(\sigma+\sigma_n(\q))(\d
R/\d\sigma)(\q,-\sigma_n(\q))}
,\eqno(4.82)
$$
or equivalently, using a generalization of the identity (C.6) to $\q\ne 0$,
together with the definition (4.70),
$$
-e_C(\q,\sigma,0,x_a)
=-2\mu_aG(\q,0)R(\q,\sigma)\sum_{n\ge
0}{R(\q,\sigma_n(\q))v_1(\q,\sigma_n(\q),x_a)
\over(\sigma+\sigma_n(\q))N_n(\q)}
.\eqno(4.83)
$$
Finally, we can derive an explicit expression for $\rho(\q,x,x_a)$,
by means of an inversion formula analogous to eq. (4.37), namely
$$
\eqalign{
\rho(\q,x,x_a)&=-2G(\q,0)\sum_{m, n\ge 0}
{R(\q,\sigma_m(\q))R(\q,\sigma_n(\q))\over\sigma_m(\q)+\sigma_n(\q)}
{v_1(\q,\sigma_m(\q),x)\over N_m(\q)}{v_1(\q,\sigma_n(\q),x_a)\over N_n(\q)}\cr
&-\sum_{n\ge 0}{
v_1(\q,\sigma_n(\q),x)v_1(\q,\sigma_n(\q),-x_a)
+v_1(\q,\sigma_n(\q),x_a)v_1(\q,\sigma_n(\q),-x)\over N_n(\q)}.
}
\eqno(4.84)
$$
This central result is manifestly symmetric under the exchange of
$x$ and $x_a$, as it should.
This symmetry property is a valuable check of the whole approach,
since both arguments $x$ and $x_a$ have played uneven roles
throughout the derivation.

We now consider the $\q\to 0$ limit of the above results, which will then
be denoted with an $S$-subscript instead of a $C$-subscript,
for the sake of consistency with the notations of section 2.
Eq. (4.83) can be recast for $\q\to 0$ as
$$
-e_S(\sigma,0,x_a)=2\mu_aP(\sigma)\left(1-{\sigma\over 3}\sum_{n\ge 1}
{\sigma_nP(\sigma_n)v_1(\sigma_n,x_a)\over(\sigma+\sigma_n)N_n}\right)
.\eqno(4.85)
$$
First, we are now able to complete the proof of the anticipated result
(4.54, 55), by inserting eq. (4.85) into eq. (4.79),
expanding the latter equation for $\q\to 0$
to the first non-trivial order as $\sigma\to 0$,
and comparing the result with the estimate
$$
a_S(\sigma,x_a)\approx-{\tau_1(\mu_a)\over\sigma}\quad(\sigma\to 0)
.\eqno(4.86)
$$
Second, the small-$\sigma$ expansion of eq. (4.85)
allows us to recover the sum rules (2.33, 37a) in the following form
$$
\int_0^\infty\rho(x,x_a)V(x)\d x=2,
\quad\int_0^\infty\rho(x,x_a)\tanh xV(x)\d x=2\nu(x_a)/3-2\tanh x_a
.\eqno(4.87)
$$
On the other hand, for large values of $\sigma=K^2$,
we can use the semi-classical estimates (B.18, 20, 22),
setting $\sigma_n=k^2$, in order to transform eqs. (4.76, 85) into
$$
\int_0^\infty\rho(x,x_a)\Ai(K^{2/3}x)x\d x\approx(2/\pi)K^{1/3}
\int_0^\infty{k^{2/3}\over k^2+K^2}\Ai(k^{2/3}x_a)\d k
.\eqno(4.88)
$$
This estimate shows that $xx_a\rho(x,x_a)$ is a homogeneous function
of its arguments with degree zero, when both of them are small,
i.e., $xx_a\rho(x,x_a)=g(x/x_a)$.
The rescaling of eq. (4.88) according to $z=K^{2/3}x=k^{2/3}x_a$ then yields,
by means of a mere identification of both integrands,
using eq. (B.25), the expression $g(z)=(3/\pi)z^{3/2}/(z^3+1)$,
implying the following scaling behavior
$$
\rho(x_a,x_b)\approx {3(x_ax_b)^{1/2}\over\pi(x_a^3+x_b^3)}
\quad(x_a,x_b\ll 1)
,\eqno(4.89)
$$
or equivalently
$$
\gamma(\mu_a,\mu_b)\approx {3(\mu_a\mu_b)^{3/2}\over\pi(\mu_a^3+\mu_b^3)}
\quad(\mu_a,\mu_b\ll 1)
.\eqno(4.90)
$$
This novel result is in contrast with the rational behavior
$\gamma(\mu_a,\mu_b)\approx\mu_a\mu_b/(\mu_a+\mu_b)$
in the case of isotropic scattering.
The law (4.90) confirms the hypothesis made in the beginning of this section,
namely that $\gamma(\mu_a,\mu_b)$ vanishes faster than linearly in
either of its arguments.
It is also worth noticing that the scaling form (4.90)
of the bistatic coefficient saturates the sum rule (2.33).

The full expression of the bistatic coefficient $\gamma(\mu_a,\mu_b)$
is obtained by taking the $\q\to 0$ limit of eq. (4.84),
using the definition (4.65).
The modes $m, n=0$ yield divergent contributions as $\q\to 0$,
which cancel out as they should,
as well as finite parts, so that we are left with the result
$$
\eqalign{
{\gamma(\mu_a,\mu_b)\over\mu_a\mu_b}
&=3\tau_0+{3\over 2}(\mu_a+\mu_b)
+2\sum_{n\ge 1}{P(\sigma_n)\over N_n}
\big[v_1(\sigma_n,x_a)+v_1(\sigma_n,x_b)\big]\cr
&-\sum_{n\ge 1}{1\over N_n}
\big[v_1(\sigma_n,x_a)v_1(\sigma_n,-x_b)
+v_1(\sigma_n,-x_a)v_1(\sigma_n,x_b)\big]\cr
&-{2\over 3}\sum_{m, n\ge 1}{\sigma_m\sigma_n\over\sigma_m+\sigma_n}
{P(\sigma_m)\over N_m}{P(\sigma_n)\over N_n}
v_1(\sigma_m,x_a)v_1(\sigma_n,x_b),\cr
}
\eqno(4.91)
$$
with $\mu_a=\tanh x_a$, $\mu_b=\tanh x_b$.

The maximal value of the bistatic coefficient,
which yields an absolute prediction for the diffuse reflected intensity
in the normal direction, reads $\gamma(1,1)=4.889703$.
This number is some 15\% above the corresponding one in the case
of isotropic scattering (see Table 2).

The shape $\gamma_C(\q)$ of the enhanced backscattering cone
at normal incidence in the regime of very anisotropic scattering
can also be derived from the general result (4.84).
Indeed, by letting $\mu_a=\mu_b=1$ in the latter formula,
and using again a generalization of the identity (C.6),
we obtain the following exact result
$$
\eqalign{
\gamma_C(\q)&=2\sum_{n\ge 0}{1\over (\d G/\d\sigma)(\q,\sigma_n(\q))}
-2G(\q,0)\sum_{m, n\ge 0}{R(\q,\sigma_m(\q))R(\q,\sigma_n(\q))
\over(\sigma_m(\q)+\sigma_n(\q))N_m(\q)N_n(\q)}
}
.\eqno(4.92)
$$
For small values of the reduced wavevector, the following behavior
$$
\gamma_C(\q)=\gamma(1,1)-\tau_1(1)^2\q/3+\O(\q^2)\quad(\q\ll 1)
\eqno(4.93)
$$
is more directly obtained from eq. (4.62).
On the other hand, for large $\q$,
the second (double) sum in eq. (4.92) is negligible,
whereas the first (simple) one can be evaluated by means of
the semi-classical estimate (B.35).
After some algebra using the identity $\lim_{z\to1}\sum_{n\ge0}(-z)^n=1/2$,
we are left with the estimate
$$
\gamma_C(\q)\approx 2\q e^{-2\q}\quad(\q\gg 1)
.\eqno(4.94)
$$

Figure 3 shows a plot of the enhancement factor $B(Q)$,
defined in eq. (2.46), against the reduced wavevector $\q=Q\sqrt{\tau^*}$,
for both isotropic and very anisotropic scattering.

A last comment is in order.
For a large but finite anisotropy,
the shape of the enhanced backscattering cone $\gamma_C(Q)$
only follows the universal scaling law (4.92)
for $\q$ of order unity, i.e.,
$Q$ of order $1/\sqrt{\tau^*}\sim\Thetarms$.
For larger values of $Q$, the backscattering cone exhibits
non-universal tails, which are expected to be generically exponentially small.

\medskip
\noindent{\bf 4.5. Extinction lengths of azimuthal excitations}

Up to this point, we have only considered samples
in the form of a slab, or a half-space,
and investigated physical quantities
which have the cylindrical symmetry around the normal to the slab,
i.e., which are independent of the azimuthal angle $\varphi$.

We now want to consider briefly the $\varphi$-dependence
of the diffuse intensity in the regime of very anisotropic scattering.
Because the sample has cylindrical symmetry,
the azimuthal dependence of physical quantities,
such as the diffuse intensity $I(\r,\n)=I(\r,\theta,\varphi)$,
can be expanded over the modes $e^{im\varphi}$,
where the algebraic integer $m$ is the azimuthal, or magnetic quantum number.
Different sectors associated with different values of $m$ are decoupled.
As already mentioned in section 2,
all the sectors contribute e.g. to the reflected intensity,
except if the incident beam is normal to the sample,
or, more generally, has itself cylindrical symmetry.
The situation is different in transmission through thick slabs,
to which only the sector $m=0$, considered so far, contributes.
The reason for this is that the intensity in the other sectors
is exponentially damped inside the sample, namely
$$
I_m\sim e^{-z/\ell_m}
,\eqno(4.95)
$$
where $\ell_m$ is the extinction length of the azimuthal excitations
in the sector $m$.

It is the purpose of this section to determine these lengths
in the limit of very anisotropic scattering.
The Legendre operator (4.8) reads
$$
\Delta_m=\cot\theta{\p\over\p\theta}+{\p^2\over\p\theta^2}
-{m^2\over\sin^2\theta}
\eqno(4.96)
$$
in the sector defined by the azimuthal integer $m$.
As a consequence, and along the lines of sections 4.3 and 4.4,
we are led to study the Schr\"odinger equation
$$
u''(\sigma,x)-\big(m^2+\sigma V(x)\big)u(\sigma,x)=0
.\eqno(4.97)
$$
The corresponding extinction length is given by
$$
\ell_m={2\ell^*\over\sigma_0^{(m)}}
,\eqno(4.98)
$$
where $\sigma_0^{(m)}$ is the smallest positive eigenvalue of eq. (4.97).
This is indeed the location of the first singularity of the Laplace
transform of the intensity.

For large values of the azimuthal integer $m$,
we can use the semi-classical analysis developed in Appendix B.
The Sommerfeld quantization formula associated with eq. (4.97) reads
$$
\int_{x_-}^{x_+}\big(-\sigma V(x)-m^2\big)^{1/2}\d x
=\int_{\mu_-}^{\mu_+}
\left(\sigma{\mu\over 1-\mu^2}-{m^2\over(1-\mu^2)^2}\right)^{1/2}\d\mu
\approx(n+1/2)\pi
.\eqno(4.99)
$$
The smallest eigenvalue $\sigma_0^{(m)}$ corresponds to
setting $n=0$ in the above formula.
In first approximation we express that the argument of the square-root
inside the $\mu$-integral in eq. (4.99) has zero as its maximal value,
so that $\mu_-=\mu_+$.
We thus obtain $\sigma_0^{(m)}\approx 3\sqrt{3}m^2/2$.
In second approximation we expand the integrand around its maximum,
which takes place for $\mu\approx\sqrt{3}/3$.
We obtain after some algebra the following next-to-leading order estimate
$$
\sigma_0^{(m)}\approx(3\sqrt{3}/2)(m^2+m\sqrt{2})
,\eqno(4.100)
$$
whence
$$
\ell_m\approx{4\ell^*\over 3\sqrt{3}(m^2+m\sqrt{2})}\quad(m\gg 1)
.\eqno(4.101)
$$

This semi-classical estimate gives accurate numbers of the whole
spectrum of extinction lengths, down to the largest one, $\ell_1$.
Indeed eq. (4.101) yields $\ell_1/\ell^*\approx 0.318861$,
whereas the exact numerical value reads $\ell_1/\ell^*=0.2829169$.

For a large but finite anisotropy,
the lengths $\ell_m$ follow the universal law (4.101)
only for $m<m^*\sim\sqrt{\tau^*}\sim 1/\Thetarms$.
For larger values of the azimuthal number,
the $\ell_m$ become non-universal numbers of order $\ell$.
This crossover is expected on physical grounds.
Indeed large azimuthal numbers $m\gg m^*$ correspond
to an angular resolution $\delta\Theta\ll\Thetarms$,
so that the details of the cross-section matter in this regime.

\medskip
\noindent{\bf 5. DISCUSSION}

In this paper we have considered several aspects
of the theory of multiple scattering of scalar waves,
with a general anisotropic cross-section.
We have limited this study to the geometry of an optically thick slab,
of thickness $L\gg\ell^*$.
Two different length scales are relevant,
namely the scattering mean free path $\ell$
and the transport mean free path $\ell^*$.
The general results of section 2 concern the dependence of various quantities
on both mean free paths.
Quantities for which only the long-distance diffusive behavior matters
scale with the transport mean free path $\ell^*$.
This is the case for e.g.
the angle-resolved transmission through a thick slab (2.42),
and especially for the thickness $z_0=\tau_0\ell^*$ of a skin layer,
i.e., the extrapolation length.
It will indeed be recalled later on that the anisotropy
of the scattering mechanism only contributes a small effect,
entirely contained in prefactors such as the constant $\tau_0$,
or the functions $\tau_1(\mu)$ and $\gamma(\mu,\mu')$.

We have also obtained an unexpected result
concerning the width of the backscattering cone,
scaling as $\Delta\theta\sim\lambda_0/\sqrt{\ell\ell^*}$ (2.60).
This prediction is in disagreement with the previously accepted
scaling law in $\lambda_0/\ell^*$ [29, 30].
The derivations of the latter result rely on the diffusion approximation.
A reason why this approximation may fail will be given in a while.
It is clearly apparent from sections 2.5, 6 that the derivation
of our scaling law for the width of the backscattering cone
is fully similar to that of the known law (2.64)
of the absorption length [2, 3], namely $L_{\rm abs}\sim\sqrt{\ell\ell^*}$.
This connection between the backscattering peak and the
absorption length is seemingly rather accidental.
As a matter of fact, such a relationship is to some extent already present
in the diffusion approximation [9].
Indeed in this framework all extinction effects,
provided they are small enough, can be coded in a single parameter,
namely the {\it mass} ${\cal M}$ such that
$$
{\cal M}^2=q^2+i\Omega+{1\over L_{\rm abs}^2}
,\eqno(5.1)
$$
where $q=Q/\ell$ is the transverse wavevector,
$L_{\rm abs}$ is the absorption length,
and $\Omega=(\omega-\omega')/D_{\rm phys}$
represents the properly dimensioned contribution of a small frequency shift
between the advanced and the retarded amplitude propagators
which build up the {\it diffuson}.
The scaling laws (2.60) and (2.64)
for the width of the cone and the absorption length
can be put together as ${\cal M}^2\sim 1/(\ell\ell^*)$.

In order to understand in heuristic terms the role played by the
intermediate length scale $\sqrt{\ell\ell^*}$,
in the case of the absorption length, we can use the following argument,
based on the random-walk picture of diffusive propagation.
If the absorption cross-section is very small,
i.e., if the albedo is very close to unity,
a photon will only be absorbed after a large number of collisions,
of order $N\sim 1/(1-a)$.
The absorption length is the typical depth reached by a photon
during $N$ steps of its random walk through the diffusive medium.
In the regime of very anisotropic scattering,
the random walk is persistent over a length $\ell^*\gg\ell$.
We thus have $L_{\rm abs}^2\sim N^*{\ell^*}^2$,
where $N^*$ is the effective number of independent steps of length $\ell^*$.
The total path length reads $N\ell\sim N^*\ell^*$,
whence $N^*\sim\ell/[\ell^*(1-a)]$,
and $L_{\rm abs}^2\sim\ell\ell^*/(1-a)$.
This is precisely eq. (2.64), up to a numerical factor.

The role of the intermediate length scale $\sqrt{\ell\ell^*}$
in the shape of the backscattering cone is more difficult
to understand at an intuitive level.
A photon entering the diffusive medium at the origin will leave
the medium at a random transverse position $\r_\perp$.
In rough terms,
the width of the cone scales as $\Delta\theta\sim\lambda_0/\Delta r_\perp$,
where $\Delta r_\perp$ is the typical transversal extent
$\vert\r_\perp\vert$ of the path of the photon in the medium.
In the case of isotropic scattering, the well-known result $\Delta Q\sim 1$
corresponds to $\Delta r_\perp\sim\ell$.
The paths which determine the width of the cone thus have a transversal extent
of order one mean free path.
As a consequence,
even though the triangular {\it shape} of the top of the cone
is determined by the longest paths, which are Brownian, and thus
correctly described by the diffusion approximation [29, 30],
the {\it width} of the cone is determined by the shortest paths,
and thus not necessarily quantitatively described
by the diffusion approximation.
If we extend the above line of reasoning to the regime of very anisotropic
scattering, we are led to consider paths with the shortest
transversal extent, such as that shown symbolically on Figure 4.
Each step makes an angle comparable to the maximal
allowed angle $\Thetarms$ with respect to
the previous one, so that the path consists of $N\sim 1/\Thetarms$ steps.
Its total length and its transversal extent scale
as $\Delta r_\perp\sim N\ell\sim\sqrt{\ell\ell^*}$.
This explains the scaling law for the width of the cone,
provided such {\it stressed paths} have the correct length scale.
The statistics of the transversal extent $r_\perp$ of persistent random walks
has been the subject of recent works [31, 32].
Unfortunately the results of numerical simulations presented there
only concern the universal diffusive fall-off
of the distribution for $r_\perp\gg\ell^*$,
so that the above heuristic idea can neither be confirmed nor infirmed.

The next question of interest concerns to what extent
physical quantities are universal, and to what extent they depend
on the particular microscopic scattering mechanism.
As far as the diffuse reflected or transmitted intensities are concerned,
the detailed structure
of the scattering mechanism only contributes a small effect,
contained in prefactors of the laws mentioned above,
such as the constant $\tau_0$,
or the functions $\tau_1(\mu)$ and $\gamma(\mu,\mu')$.
Two situations allow for a more detailed investigation of these matters.

\noindent (i)
The regime of a large index mismatch,
where the boundaries of the sample almost act as perfect mirrors,
is considered in section 3.
Our main result (3.9) shows that the physical quantities considered
do not depend {\it at all} on the scattering cross-section in this regime.
This can be easily understood as follows.
Since the thickness $z_0\approx 4\ell^*/(3\T)$ of a skin layer is very large,
the radiation undergoes many scattering events near the boundaries
before it leaves the medium,
so that the details of every single scattering event are washed out.

\noindent (ii)
In the absence of internal reflections,
we have considered in detail the regime of very anisotropic scattering.
Our exact treatment, based on the expansion (4.7),
is expected to be valid in the regime $\Thetarms\ll 1$
of a broad {\it universality class} of phase functions.
Although this universality class cannot be easily characterized,
we can assert that it contains at least the phase functions scaling as
$$
p(\Theta)\approx\Phi\big(\Theta/\Thetarms\big)
,\eqno(5.2)
$$
such that the scaling function $\Phi$ has a finite second moment.
This restrictive definition does not encompass a priori the Lorentzian-squared
phase function (4.4), which has a logarithmically divergent second moment,
as already mentioned in section 4.1.
The same remark holds for the so-called Henyey-Greenstein phase function
$$
p(\Theta)={1-g^2\over(1-2g\cos\Theta+g^2)^{3/2}}
,\eqno(5.3)
$$
often used in numerical investigations [3, 14],
for which the second-moment integral is linearly divergent.
In section 4 we have presented an exact analytical treatment
of universal aspects of very anisotropic scattering,
yielding to the results (4.48, 55, 91, 92) for the observables
pertaining to thick slabs.
These results are not entirely explicit,
since they are given in terms of the eigenvalues and eigenfunctions of
one-dimensional Schr\"odinger equations,
which are accessible both numerically,
via the partial-wave expansion of Appendix A,
and analytically in the limit of large quantum numbers,
via the semi-classical analysis of Appendix B.
The exact solution derived there is nevertheless an analogue
in the regime of very anisotropic scattering
of the exact solution in the case of isotropic scattering,
which dates back to Chandrasekhar [1].

The discussion of the dependence of physical quantities
on the details of the scattering mechanism is summarized in Table 2,
where we compare the numerical values of the dimensionless
absolute prefactors of five characteristic quantities,
for isotropic scattering and for very anisotropic scattering.
The relative differences, shown in the last row, are very small in most cases.
Some other quantities, such as the shape of the enhanced backscattering cone,
or the spectrum of extinction lengths of the azimuthal excitations,
exhibit universal behavior in the very anisotropic regime
only in a limited range, corresponding to a low enough angular resolution
$(\delta\Theta\gg\Thetarms)$,
so that the details of the scattering cross-section do not matter.

Finally, we can compare our universal results
in the very anisotropic scattering regime,
for some of the quantities listed in Table 2,
with the outcomes of numerical approaches.
Van de Hulst [3, 14] has investigated in a systematic way
the dependence of various quantities on anisotropy,
for several commonly used phenomenological forms of the phase function,
including especially the Henyey-Greenstein phase function (5.3).
The data on the skin-layer thickness reported in ref. [3]
show that, as a function of anisotropy,
$\tau_0$ varies from 0.7104 (isotropic scattering)
to 0.7150 (moderate anisotropy), passing a minimum of 0.7092 (weak anisotropy).
The trend shown by these data suggests that our universal value 0.718211
is actually an absolute upper bound for $\tau_0$.
Numerical data concerning $\tau_1(1)$ is also available.
Van de Hulst [14] has extrapolated two series of data,
concerning the Henyey-Greenstein phase function (5.3),
which admit a common limit for very anisotropic scattering $(g\to 1)$.
According to the analysis of section 4,
this limit reads in our language $\tau_1(1)/4=1.284645$,
whereas ref. [14] gives the two slightly different estimates
$1.273\pm 0.002$ and $1.274\pm 0.007$.
The agreement is satisfactory,
although it cannot be entirely excluded that the observed 0.8\%
relative difference can be a small but genuine nonuniversality effect.
Indeed, as mentioned above, the Henyey-Greenstein phase function (5.3)
might not belong to the universality class where our approach holds true.
The same remark applies to a less complete set of data [14] concerning
the intensity $\gamma(1,1)$ of reflected light at normal incidence.

\medskip
\noindent{\bf Acknowledgments}

We are pleased to thank H. van de Hulst for his interest in this work,
for his useful comments,
and for having communicated to us his recent unpublished notes [14].
We thank J.P. Bouchaud, B. van Tiggelen, and A. Voros
for stimulating discussions.
J.P. Bouchaud and H. van de Hulst are also acknowledged
for a careful reading of the manuscript.
Th.M.N. acknowledges SPhT (Saclay) for their hospitality.
His research work has been made possible by support by the
Royal Netherlands Academy of Arts and Sciences (KNAW).
E.A. acknowledges the University of Amsterdam for their hospitality.

\vfill\eject
\noindent{\bf Appendix A. Partial-wave expansions}

In this Appendix we describe a numerical algorithm
based on a partial-wave expansion,
that we have used to determine the eigenvalues and eigenfunctions
of the Schr\"odinger equations involved in section 4.

We first consider the Schr\"odinger equation (4.22).
Going back to the $\mu$-variable, this equation reads
$$
(\Delta_0-\sigma\mu)v_1(\sigma,\mu)=0
,\eqno({\rm A}.1)
$$
where $\Delta_0$ is the Legendre operator in the $\varphi$-independent sector,
defined in eq. (4.10).
It is natural to expand the function $v_1(\sigma,\mu)$
in the Legendre polynomials
$$
v_1(\sigma,\mu)=\sum_{\ell\ge 0}a_\ell(\sigma)P_\ell(\mu)
.\eqno({\rm A}.2)
$$
Indeed these polynomials are eigenfunctions of $\Delta_0$, namely
$$
\Delta_0P_\ell=-\ell(\ell+1)P_\ell
,\eqno({\rm A}.3)
$$
and the product $\mu P_\ell(\mu)$ has the following expression
$$
(2\ell+1)\mu P_\ell(\mu)=(\ell+1)P_{\ell+1}(\mu)+\ell P_{\ell-1}(\mu)
,\eqno({\rm A}.4)
$$
so that eq. (A.1) amounts to the following three-term recursion relation
$$
\ell(\ell+1)a_\ell
+\sigma\left({\ell+1\over2\ell+3}a_{\ell+1}
+{\ell\over2\ell-1}a_{\ell-1}\right)=0
.\eqno({\rm A}.5)
$$

When $\sigma$ is one of the eigenvalues $\sigma_n$,
eq. (A.5) has an acceptable solution $\{a_\ell(\sigma)\}$,
decaying to zero for large $\ell$.
The quantities needed in section 4 can then be evaluated as follows.
The normalization condition (4.24) becomes
$$
\sum_{\ell\ge 0}a_\ell(\sigma_n)=1
,\eqno({\rm A}.6)
$$
since $P_\ell(1)=1$.
The squared norms $N_n$ of the eigenfunctions read
$$
N_n=4\sum_{\ell\ge 0}{\ell+1\over(2\ell+1)(2\ell+3)}
a_\ell(\sigma_n)a_{\ell+1}(\sigma_n)
,\eqno({\rm A}.7)
$$
as a consequence of the normalization of the Legendre polynomials
$$
\int_{-1}^1{\d\mu\over 2}P_k(\mu)P_\ell(\mu)={\delta_{k,\ell}\over 2\ell+1}
.\eqno({\rm A}.8)
$$
Finally, the non-trivial mixed Wronskians $F(\pm\sigma_n)$ read
$$
F(-\sigma_n)={1\over F(\sigma_n)}
=v_1(\sigma_n,\mu=-1)=\sum_{\ell\ge 0}(-1)^\ell a_\ell(\sigma_n)
,\eqno({\rm A}.9)
$$
since $P_\ell(-1)=(-1)^\ell$.

We now consider the $\q$-dependent Schr\"odinger equation (4.66), namely
$$
\big(\Delta_0-\sigma\mu+\q^2(\mu^2-1)\big)v_1(\q,\sigma,\mu)=0
.\eqno({\rm A}.10)
$$
We again expand the wavefunction over the Legendre polynomials
$$
v_1(\q,\sigma,\mu)=\sum_{\ell\ge 0}a_\ell(\q,\sigma)P_\ell(\mu)
.\eqno({\rm A}.11)
$$
By iterating eq. (A.4) twice, we obtain the following five-term recursion
$$
\eqalign{
\ell(\ell+1)a_\ell
&+\sigma
\left({\ell+1\over2\ell+3}a_{\ell+1}+{\ell\over2\ell-1}a_{\ell-1}\right)\cr
&-\q^2\left({\ell(\ell-1)\over(2\ell-1)(2\ell-3)}a_{\ell-2}
+{2(1-\ell-\ell^2)\over(2\ell-1)(2\ell+3)}a_\ell
+{(\ell+1)(\ell+2)\over(2\ell+3)(2\ell+5)}a_{\ell+2}\right)=0.
}
\eqno({\rm A}.12)
$$
Eqs. (A.6, 7, 9) still hold true.

We finally consider the wave equation
$$
(\Delta-\sigma\mu)v_1(\sigma,\mu,\varphi)=0
,\eqno({\rm A}.13)
$$
where $\Delta$ is the full Legendre operator, defined in eq. (4.8).
Since the potential does not involve the azimuthal angle $\varphi$ explicitly,
we look for a solution $v_1$ proportional to $e^{im\varphi}$,
with $m\ge 0$ being an integer.
It is now natural to expand the function $v_1(\sigma,\mu,\varphi)$
in the spherical harmonic functions, which we choose to write as follows
$$
v_1(\sigma,\mu,\varphi)=e^{im\varphi}
\sum_{\ell\ge m}a_{\ell,m}(\sigma)P_\ell^m(\mu)
,\eqno({\rm A}.14)
$$
in terms of the Legendre functions $P_\ell^m(\mu)$, which obey
$$
\Delta\big(P_\ell^m(\mu)e^{im\varphi}\big)
=-\ell(\ell+1)P_\ell^m(\mu)e^{im\varphi}
,\eqno({\rm A}.15)
$$
and the product $\mu P_\ell^m(\mu)$ has the following expression
$$
(2\ell+1)\mu P_\ell^m(\mu)
=(\ell+1-m)P_{\ell+1}^m(\mu)+(\ell+m)P_{\ell-1}^m(\mu)
,\eqno({\rm A}.16)
$$
so that eq. (A.13) amounts to the three-term recursion relation
$$
\ell(\ell+1)a_{\ell,m}
+\sigma\left({\ell+m+1\over2\ell+3}a_{\ell+1,m}
+{\ell-m\over2\ell-1}a_{\ell-1,m}\right)=0
.\eqno({\rm A}.17)
$$
The recursion equations (A.5, 12, 17) are easily implemented numerically.

\vfill\eject
\noindent{\bf Appendix B. Semi-classical analysis}

The outcomes of the semi-classical analysis presented in this Appendix
are used at various places in section 4.
We first consider the Schr\"odinger equation (4.22).
For large values of the complex parameter $\sigma$,
we look for rapidly varying solutions of the form
$$
u\approx\Phi(x)^{-1/2}\exp\int^x\Phi(y)\d y
,\eqno({\rm B}.1)
$$
with
$$
\Phi(x)^2=\sigma V(x)
.\eqno({\rm B}.2)
$$
This approach is analogous to the celebrated W.K.B. approximation
in Quantum Mechanics,
since the condition $\sigma\gg 1$ is equivalent to $\hbar$ being small.

Because the potential $V(x)$ is odd,
we can restrict the analysis to the domain $\Re\sigma>0$.
We introduce the notation
$$
K=\sqrt{\sigma}
.\eqno({\rm B}.3)
$$

Eq. (B.2) has real solutions for $x>0$.
We set
$$
p(x)=\sqrt{V(x)}\quad(x>0)
,\eqno({\rm B}.4)
$$
so that $\Phi(x)=Kp(x)$.
Eq. (B.1) thus yields the basis of functions
$$
u_{\pm}(x)\approx{1\over\big(Kp(x)\big)^{1/2}}\exp\big(\pm KI(x)\big)
,\eqno({\rm B}.5)
$$
with
$$
I(x)=\int_x^\infty p(y)\d y\quad(x>0)
.\eqno({\rm B}.6)
$$
The above functions $u_\pm(x)$ are exponentially blowing up or decaying.
The domains $x>0$ and $\sigma>0$ (and similarly $x<0$ and $\sigma<0$)
are said to be {\it classically forbidden}.

On the other hand, for $x<0$, eq. (B.2) has imaginary solutions.
We set
$$
q(x)=\sqrt{-V(x)}\quad(x<0)
,\eqno({\rm B}.7)
$$
so that $\Phi(x)=iKq(x)$.
Eq. (B.1) thus yields the basis of functions
$$
u_{\pm}(x)\approx{1\over\big(Kq(x)\big)^{1/2}}\exp\big(\pm iKI(x)\big)
,\eqno({\rm B}.8)
$$
with
$$
I(x)=\int_{-\infty}^{-x} q(y)\d y\quad(x<0)
.\eqno({\rm B}.9)
$$
The above functions $u_\pm(x)$ are oscillating and bounded,
up to the prefactor in $q(x)^{-1/2}$.
The domains $x<0$ and $\sigma>0$ (and similarly $x>0$ and $\sigma<0$)
are said to be {\it classically allowed}.

The difficulty of the semi-classical analysis comes from the existence
of three {\it turning points}, namely $x=0$ and $x\to\pm\infty$,
where the momentum variable $p(x)$ or $q(x)$ vanishes.
The estimates (B.5, 8) loose their meaning in the vicinity
of the turning points,
where a more careful analysis is required, to be presented now.

We first investigate the basis of functions $\{u_1,u_2\}$ for $x<0$.
For $x\to -\infty$ and $K\to\infty$,
the Schr\"odinger equation (4.22) assumes the simpler form
$$
u''+(2Ke^x)^2u=0
.\eqno({\rm B}.10)
$$
A basis of solutions to this equation is given by the Bessel functions
$J_0(z)$ and $N_0(z)$, with $z=2Ke^x$ being a scaling variable.
The boundary conditions (4.23) have to match the known small-$z$ behavior of
the Bessel functions, whence
$$
\eqalign{
u_1(x)&\approx J_0(2Ke^x),\cr
u_2(x)&\approx(\pi/2)N_0(2Ke^x)-(\ln K+\euler)J_0(2Ke^x)\quad(x\to -\infty),
}
\eqno({\rm B}.11)
$$
where $\euler$ denotes Euler's constant.
The known large-$z$ behavior of the Bessel functions fixes
the amplitudes of the integrals in eq. (B.8) for $u_1$ and $u_2$, namely
$$
\eqalign{
u_1(x)&\approx\left({2\over\pi Kq(x)}\right)^{1/2}
\cos\big(KI(x)-\pi/4\big),\cr
u_2(x)&\approx\left({2\over\pi Kq(x)}\right)^{1/2}
\left[(\pi/2)\sin\big(KI(x)-\pi/4\big)
-(\ln K+\euler)\cos\big(KI(x)-\pi/4\big)\right].\cr
}
\eqno({\rm B}.12)
$$
On the other hand, for $x\to 0$ and $K\to\infty$,
the Schr\"odinger equation (4.22) assumes the simpler form
$$
u''+K^2xu=0
,\eqno({\rm B}.13)
$$
which is equivalent to Airy's equation.
A basis of solutions is given by the Airy functions
$\Ai(z)$ and $\Bi(z)$, with $z=K^{2/3}x$ being again a scaling variable.
The known behavior for $z\to -\infty$ of the Airy functions
has to match eq. (B.12), whence
$$
\eqalign{
u_1(x)\approx 2^{1/2}K^{-1/3}
&\left[\sin(KI)\Ai(K^{2/3}x)+\cos(KI)\Bi(K^{2/3}x)\right],\cr
u_2(x)\approx 2^{1/2}K^{-1/3}
&\left\{
(\pi/2)\left[-\cos(KI)\Ai(K^{2/3}x)+\sin(KI)\Bi(K^{2/3}x)\right]\right.
\cr
&\left.
-(\ln K+\euler)\left[\sin(KI)\Ai(K^{2/3}x)+\cos(KI)\Bi(K^{2/3}x)\right]\right\}
,\cr}
\eqno({\rm B}.14)
$$
for $x\to 0$, with
$$
I=I(0)=\int_0^\infty p(x)\d x
=\int_0^1\d\mu\left({\mu\over 1-\mu^2}\right)^{1/2}
={\sqrt{\pi}\over 2}{\Gamma(3/4)\over\Gamma(5/4)}=1.198140
,\eqno({\rm B}.15)
$$
this definition being consistent with eqs. (B.6, 9).

We now investigate in a similar way the functions $\{v_1,v_2\}$ for $x>0$.
For $x\to\infty$ and $K\to\infty$,
a basis of solutions is given by the modified Bessel functions
$I_0(z)$ and $K_0(z)$, with $z=2Ke^{-x}$.
The boundary conditions (4.24) have to match the known small-$z$ behavior of
the Bessel functions, whence
$$
\eqalign{
v_1&\approx I_0(2Ke^{-x}),\cr
v_2&\approx K_0(2Ke^{-x})+(\ln K+\euler)I_0(2Ke^{-x})\quad(x\to\infty).
}
\eqno({\rm B}.16)
$$
The known large-$z$ behavior of the Bessel functions only fixes
the amplitude of the solution $u_+$ of eq. (B.5) for $v_1$ and $v_2$, namely
$$
\eqalign{
v_1(x)&\approx\left({1\over 2\pi Kp(x)}\right)^{1/2}
e^{KI(x)},\cr
v_2(x)&\approx(\ln K+\euler)v_1(x)\quad(x>0).
}
\eqno({\rm B}.17)
$$
Finally, the known behavior of the Airy functions as $z\to\infty$ yields
$$
\eqalign{
v_1(x)&\approx 2^{1/2}K^{-1/3}\Ai(K^{2/3}x)e^{KI},\cr
v_2(x)&\approx(\ln K+\euler)v_1(x)\quad(x\to 0).
}
\eqno({\rm B}.18)
$$

The above expressions (B.14, 18) of both bases of solutions
as $x\to 0$ and $K\to\infty$
allow us to derive the following semi-classical estimates
for the elements of the transfer matrix introduced in eq. (4.28)
$$
\eqalign{
F(\sigma)&\approx\big[\sin(KI)-(2/\pi)(\ln K+\euler)\cos(KI)\big]e^{KI},\cr
G(\sigma)&\approx-(2/\pi)\cos(KI)e^{KI},\cr
H(\sigma)&\approx(\ln K+\euler)F(\sigma),\cr
F(-\sigma)&\approx(\ln K+\euler)G(\sigma)
\quad(\sigma=K^2\to\infty).\cr
}
\eqno({\rm B}.19)
$$

The estimate for $G(\sigma)$ directly yields the following
expressions for its factors $P(\pm\sigma)$, defined in eq. (4.39)
$$
\eqalign{
P(\sigma)&\approx (3/\pi)^{1/2}{e^{KI}\over K^2},\cr
P(-\sigma)&\approx -2(3/\pi)^{1/2}{\cos(KI)\over K^2}
\quad(\sigma=K^2\to\infty).\cr
}
\eqno({\rm B}.20)
$$

We also obtain from eq. (B.19) an estimate of the eigenvalues $\sigma_n=K_n^2$,
which are the zeroes of $G(\sigma)$, in the form
$$
K_n\approx\left(n+1/2\right){\pi\over I}\quad(n\gg 1)
.\eqno({\rm B}.21)
$$
This semi-classical formula gives a very accurate description
of the whole spectrum of the Schr\"odinger equation (4.22).
Indeed the relative error is maximal for the first non-zero eigenvalue,
for which eq. (B.21) predicts $K_1\approx 3.933086$,
i.e., some $3.2\%$ above the exact numerical value $K_1=3.811562$.

The semi-classical expression of the squared norms $N_n$ can be evaluated
by inserting the estimates (B.19) into the identity (C.6).
We thus obtain
$$
N_n\approx -{I\over\pi K_n}e^{2K_nI}\quad(n\gg 1)
.\eqno({\rm B}.22)
$$

In section 4 we also need the expression of
the Mellin transform of the Airy function $\Ai(x)$, namely
$$
m(s)=\int_0^\infty x^s\Ai(x)\d x\quad(\Re s>-1)
,\eqno({\rm B}.23)
$$
which we have not found in standard handbooks.
The Airy equation implies the functional equation
$$
m(s+3)=(s+1)(s+2)m(s)
,\eqno({\rm B}.24)
$$
whose correctly normalized solution is
$$
m(s)=3^{-s/3}{\Gamma(s)\over\Gamma(s/3)}
.\eqno({\rm B}.25)
$$

We now generalize the above approach
to the $\q$-dependent Schr\"odinger equation (4.66).
In a first step we only consider the spheroidal equation (4.72),
corresponding to $\sigma=0$.
The semi-classical regime corresponds to $\vert\q\vert\gg 1$;
the following two cases have to be considered separately.

First, for $\q$ real, the analogues of eqs. (B.11, 12) read
$$
\eqalign{
u_1(\q,x)&\approx I_0\big(2\q e^{2x}\big)\quad(\q\to\infty,\, x\to-\infty),\cr
u_1(\q,x)&\approx {1\over\big(\pi\q(1-\tanh^2 x)\big)^{1/2}}
e^{\q(1+\tanh x)}\quad(x<0).
}
\eqno({\rm B}.26)
$$
Since the potential $W(x)$ is even, we have $v_1(\q,x)=u_1(\q,-x)$, whence
$$
G(\q,0)=2u_1(\q,0)u'_1(\q,0)\approx{2\over\pi}e^{2\q}\quad(\q\to\infty)
.\eqno({\rm B}.27)
$$

Second, for $\q$ imaginary, we set $\q=i\xi$.
The analogues of eqs. (B.11, 12) read
$$
\eqalign{
u_1(\q,x)&\approx J_0\big(2\xi e^{2x}\big)\quad(\xi\to\infty,\,x\to-\infty),\cr
u_1(\q,x)&\approx {2\over\big(\pi\xi(1-\tanh^2 x)\big)^{1/2}}
\cos\big(\xi(1+\tanh x)-\pi/4\big)\quad(x<0).
}
\eqno({\rm B}.28)
$$
The spheroidal equation (4.72) has an eigenvalue whenever
either $u_1(\q,x)$ or its derivative with respect to
$x$ vanishes for $x=0$.
We thus obtain eigenvalues of the form $\q=\pm i\xi_n$,
as well as the semi-classical estimate
$$
\xi_n\approx (n+1/2)\pi/2\quad(n\gg 1)
.\eqno({\rm B}.29)
$$

In the general case where both $\sigma$ and $\q$ are non-zero,
only the spectrum of eq. (4.66) will be needed.
Hence we can content ourselves with the Sommerfeld quantization formula.
A similar treatment is used in section 4.5 for the full Legendre operator.

For the Schr\"odinger equation (4.22) the Sommerfeld formula reads
$$
\int_{-\infty}^0\big(-\sigma V(x)\big)^{1/2}\d x
\approx\left(n+1/2\right)\pi
,\eqno({\rm B}.30)
$$
yielding again the estimate $\sigma\approx\pm K_n^2$,
with $K_n$ as in eq. (B.21).

For the $\q$-dependent Schr\"odinger equation (4.66),
the Sommerfeld formula reads
$$
\int_{x_-}^{x_+}\big(-\sigma V(x)-\q^2 W(x)\big)^{1/2}\d x
=\int_{\mu_-}^{\mu_+}\left(\sigma{\mu\over 1-\mu^2}-\q^2\right)^{1/2}\d\mu
\approx\left(n+1/2\right)\pi
,\eqno({\rm B}.31)
$$
the integral being extended over the classically allowed domain,
where the square-root is real.

This implicit equation (B.31) for the semi-classical estimate
of the eigenvalues $\pm\sigma_n(\q)$ can be investigated in several
limiting cases of interest.
For small values of $\q$, the integrand can be expanded
in a straightforward way.
We thus obtain
$$
\sigma_n(\q)^{1/2}\approx K_n(\q)=\left(n+1/2\right){\pi\over I}
\left(1+{2\q^2\over 3\pi(n+1/2)^2}+\cdots\right)\quad(n\gg 1,\,\q\ll 1)
.\eqno({\rm B}.32)
$$
This expression confirms the general result (4.67).
On the other hand, for large $\q$,
the allowed domain shrinks down to a small vicinity of $\mu=1$.
We obtain the spectrum of an effective harmonic oscillator, namely
$$
\sigma_n(\q)\approx 2(2n+1)\q\quad(\q\gg 1)
.\eqno({\rm B}.33)
$$
This formula holds for all values of $n$, down to $n=0$.
The definition (4.71) thus yields
$$
R(\q,\sigma)R(\q,-\sigma)\approx\prod_{n\ge 0}
\left(1-{\sigma^2\over\big(2(2n+1)\q\big)^2}\right)=\cos{\pi\sigma\over 4\q}
\quad(\q\gg 1)
.\eqno({\rm B}.34)
$$
Finally, putting together the estimates (B.27) and (B.34),
and using the definition (4.70), we obtain the following asymptotic formula
$$
G(\q,\sigma)\approx{2\over\pi}e^{2\q}\cos{\pi\sigma\over 4\q}
\quad(\q\gg 1)
.\eqno({\rm B}.35)
$$

\vfill\eject
\noindent{\bf Appendix C. Useful identities on the mixed Wronskians}

In this Appendix, we derive the identity (C.6) used in section 4,
and more generally we give alternative expressions
for the derivatives with respect to
the spectral variable $\sigma$ of the mixed
Wronskians $F(\sigma)$, $G(\sigma)$, and $H(\sigma)$,
which enter the transfer matrix (4.28).

To do so, we start by considering the derivatives
$$
U_\alpha(\sigma,x)={\p u_\alpha(\sigma,x)\over\p\sigma}\quad(\alpha=1,2)
,\eqno({\rm C}.1)
$$
which obey the inhomogeneous Schr\"odinger equation
$$
U_\alpha''(\sigma,x)-\sigma V(x)U_\alpha(\sigma,x)=V(x)u_\alpha(\sigma,x)
\quad(\alpha=1,2)
.\eqno({\rm C}.2)
$$
These equations can be solved explicitly by {\it varying the constants},
along the lines of sections 4.3 and 4.4.
We thus obtain
$$
\eqalign{
U_1(\sigma,x)&=-u_1(\sigma,x)\int_{-\infty}^xu_1(\sigma,y)u_2(\sigma,y)V(y)\d y
+u_2(\sigma,x)\int_{-\infty}^xu_1^2(\sigma,y)V(y)\d y,\cr
U_2(\sigma,x)&=-u_1(\sigma,x)\int_{-\infty}^xu_2^2(\sigma,y)V(y)\d y
+u_2(\sigma,x)\int_{-\infty}^xu_1(\sigma,y)u_2(\sigma,y)V(y)\d y.}
\eqno({\rm C}.3)
$$

By taking the $x\to\infty$ limit of the above expressions,
and using the asymptotic behavior (4.29), we get the following expressions
$$
\eqalign{
{\d F(\sigma)\over\d\sigma}&=
G(\sigma)N_{22}(\sigma)+F(\sigma)N_{12}(\sigma),\cr
{\d G(\sigma)\over\d\sigma}&=
-F(\sigma)N_{11}(\sigma)-G(\sigma)N_{12}(\sigma),\cr
{\d H(\sigma)\over\d\sigma}&=
-F(\sigma)N_{11}(\sigma)-G(\sigma)N_{12}(\sigma),\cr
{\d F(-\sigma)\over\d\sigma}&=
-F(-\sigma)N_{12}(\sigma)-H(\sigma)N_{11}(\sigma),
}
\eqno({\rm C}.4)
$$
with the definition
$$
N_{\alpha\beta}(\sigma)=\int_{-\infty}^\infty
u_\alpha(\sigma,x)u_\beta(\sigma,x)V(x)\d x\quad(\alpha,\beta=1,2)
.\eqno({\rm C}.5)
$$

On the spectrum, i.e., for $\sigma=\sigma_n$, we have $G(\sigma_n)=0$,
by definition.
Furthermore eq. (4.33) implies $N_{11}(\sigma_n)=N_n/F^2(\sigma_n)$,
whence the identity
$$
\left({\d G(\sigma)\over\d\sigma}\right)_{\sigma=\sigma_n}
=-{N_n\over F(\sigma_n)}=-N_nF(-\sigma_n)
,\eqno({\rm C}.6)
$$
with $N_n$ being the squared norm of the eigenfunction $v_1(\sigma_n,x)$,
defined in eq. (4.34).
This result is very general;
it also holds for the $\q$-dependent Schr\"odinger equation (4.66).

\vfill\eject
{\parindent 0pt
{\bf REFERENCES}
\bigskip

[1] S. Chandrasekhar, {\it Radiative Transfer} (Dover, New-York, 1960).

[2] A. Ishimaru, {\it Wave Propagation and Scattering in Random Media}, in 2
volumes (Academic, New-York, 1978).

[3] H.C. van de Hulst, {\it Multiple Light Scattering}, in 2 volumes (Academic,
New-York, 1980).

[4] M.P. van Albada and A. Lagendijk, Phys. Rev. Lett. {\bf 55}, 2692 (1985);
P.E. Wolf and G. Maret, Phys. Rev. Lett. {\bf 55}, 2696 (1985).

[5] M.C.W. van Rossum, Th.M. Nieuwenhuizen, and R. Vlaming, Phys. Rev. E (1995)
(to appear).

[6] P.A. Lee, A.D. Stone, and H. Fukuyama, Phys. Rev. B {\bf 35}, 1039 (1987).

[7] P.N. den Outer, Th.M. Nieuwenhuizen, and A. Lagendijk, J. Opt. Soc. Am. A
{\bf 10}, 1209 (1993).

[8] A.A. Abrikosov, L.P. Gorkov, and I.E. Dzyaloshinski, {\it Methods of
Quantum Field Theory in Statistical Physics} (Prentice-Hall, Englewood Cliffs,
1963).

[9] Th.M. Nieuwenhuizen, {\it Veelvoudige verstrooing van golven}, lecture
notes in Dutch (University of Amsterdam, 1993, unpublished).

[10] G. Placzek and W. Seidel, Phys. Rev. {\bf 72}, 550 (1947).

[11] E.E. Gorodnichev, S.L. Dudarev, and D.B. Rogozkin, Phys. Lett. A {\bf
144}, 48 (1990).

[12] Th.M. Nieuwenhuizen and J.M. Luck, Phys. Rev. E {\bf 48}, 569 (1993).

[13] V.D. Ozrin, Phys. Lett. A {\bf 162}, 341 (1992).

[14] H.C. van de Hulst, unpublished preprints and notes.

[15] J.F. de Boer, M.C.W. van Rossum, M.P. van Albada, Th.M. Nieuwenhuizen, and
A. Lagendijk, Phys. Rev. Lett. {\bf 73}, 2567 (1994).

[16] Th.M. Nieuwenhuizen and M.C.W. van Rossum, Phys. Rev. Lett. (1995) (to
appear).

[17] M.J. Stephen and G. Cwilich, Phys. Rev. B {\bf 34}, 7564 (1986).

[18] M.I. Mishchenko, Phys. Rev. B {\bf 44}, 12597 (1991).

[19] A.A. Golubentsev, Sov. Phys. JETP {\bf 59}, 26 (1984).

[20] F.C. MacKintosh and S. John, Phys. Rev. B {\bf 37}, 1884 (1988).

[21 F.A. Erbacher, R. Lenke, and G. Maret, Europhys. Lett. {\bf 21}, 551
(1993).

[22] A. Lagendijk, R. Vreeker, and P. de Vries, Phys. Lett. {\bf 136}, 81
(1989).

[23] J.X. Zhu, D.J. Pine, and D.A. Weitz, Phys. Rev. A {\bf 44}, 3948 (1991).

[24] R.R. Alvano (ed.), O.S.A. Proceedings on Advances in Optical Imaging and
Photon Migration, vol. {\bf 21} (1994).

[25] H.C. van de Hulst, {\it Light Scattering by Small Particles} (Wiley,
New-York, 1957).

[26] K.F. Freed, Adv. Chem. Phys. {\bf 22}, 1 (1972).

[27] A. Erd\'elyi (ed.), {\it Higher Transcendental Functions}, vol. III
(McGraw-Hill, New-York, 1955).

[28] M. Warner, J.M.F. Gunn, and A.B. Baumg\"artner, J. Phys. A {\bf 18}, 3007
(1985).

[29] E. Akkermans and R. Maynard, J. Physique (France) Lett. {\bf 46}, L 1045
(1985); E. Akkermans, P.E. Wolf, R. Maynard, and G. Maret, J. Physique (France)
{\bf 49}, 77 (1988).

[30] Yu.N. Barabanenkov and V.D. Ozrin, Sov. Phys. JETP {\bf 67}, 1117 and 2175
(1988).

[31] M.S. Patterson, B. Chance, and B.C. Wilson, Appl. Opt. {\bf 28}, 2331
(1989).

[32] A.H. Gandjbakhche, R.F. Bonner, and R. Nossal, J. Stat. Phys. {\bf 69}, 35
(1992); A.H. Gandjbakhche, R. Nossal, and R.F. Bonner, Appl. Opt. {\bf 32}, 504
(1993).

\vfill\eject
{\bf CAPTIONS FOR TABLES AND FIGURES}
\bigskip
{\bf Figure 1:} Laws of geometrical optics for the refraction of light
by a large dielectric sphere.

{\bf Figure 2:} Plot of exact expressions for $\tau_1(\mu)$
in the absence of internal reflections.
Dashed line: isotropic scattering, after ref. [12].
Full line: very anisotropic scattering (this work).

{\bf Figure 3:} Plot of exact expressions for the enhancement factor
$B(Q)$ of backscattering cone at normal incidence,
in the absence of internal reflections,
against $\q=Q\sqrt{\tau^*}=k_1\sqrt{\ell\ell^*}\theta$.
Dashed line: isotropic scattering, after ref. [12].
Full line: very anisotropic scattering (this work).

{\bf Figure 4:} Schematic plot of a stressed path yielding the shortest
value of the transversal extent $r_\perp$ compatible with the opening
angle $\Thetarms$.

\bigskip

{\bf Table 1:}
Conventions and notations for kinematic and other useful quantities.

{\bf Table 2:}
Comparison of the numerical values
of various quantities of interest from the exact solutions
in the absence of internal reflections.
First row: isotropic scattering, after ref. [12];
second row: very anisotropic scattering (this work).
$\tau_0\ell^*$ is the thickness of a skin layer;
$\tau_1(1)$ and $\gamma(1,1)$ respectively yield the
transmitted and reflected intensities in the normal direction;
$B(0)$ is the peak value of the enhancement factor
at the top of the backscattering cone;
$\Delta\q$ is the width of the backscattering cone,
in units of $\q=Q\sqrt{\tau^*}=k_1\sqrt{\ell\ell^*}\theta$.
The third row gives the relative difference of the second case
with respect to the first one.
}
\vfill\eject

\centerline {\bf Table 1}
\smallskip
$$\vbox{\init\halign to 16truecm
{\strut#&\vrule#\tabskip=1em plus 2em&
\hfil$#$\hfil&\vrule#&\hfil$#$\hfil&\vrule#&\hfil$#$\hfil&
\vrule#\tabskip 0pt\crr
&&\ &&\ &&\ &\cr
&&\ &&\hbox{outside medium} &&\hbox{inside medium} &\cr
&&\ &&\ &&\ &\crr
&&\ &&\ &&\ &\cr
&&\hbox{optical index} &&n_1 &&n_0=m n_1 &\cr
&&\ &&\ &&\ &\crr
&&\ &&\ &&\ &\cr
&&\hbox{wavenumber} &&k_1=n_1\omega/c=2\pi/\lambda_1
&&k_0=n_0\omega/c=mk_1=2\pi/\lambda_0 &\cr
&&\ &&\ &&\ &\crr
&&\ &&\ &&\ &\cr
&&\hbox{incidence angle} &&\theta &&\theta' &\cr
&&\ &&\ &&\ &\crr
&&\ &&\ &&\ &\cr
&&\hbox{parallel wavevector} &&
\eqalign{p &=k_1\cos\theta\cr &=k_0\sqrt{\mu^2-1+1/m^2}} &&
\eqalign{P &=k_0\cos\theta'\cr &=k_0\mu} &\cr
&&\ &&\ &&\ &\crr
&&\ &&\ &&\ &\cr
&&
\matrix{\hbox{total reflection}\cr\hbox{condition}} &&
\matrix{m<1\ \ \hbox{and}\ \ \sin\theta>m\cr (\hbox{i.e.}\ \ P\ \
\hbox{imaginary})} &&
\matrix{m>1\ \ \hbox{and}\ \ \sin\theta'>1/m \cr (\hbox{i.e.}\ \ p\ \
\hbox{imaginary})} &\cr
&&\ &&\ &&\ &\crr
}}$$
$$\vbox{\init\halign to 16truecm
{\strut#&\vrule#\tabskip=1em plus 2em&
\hfil$#$\hfil&\vrule#&\hfil$#$\hfil&
\vrule#\tabskip 0pt\crr
&&\ &&\ &\cr
&&\matrix{{\rm transverse}\cr{\rm wavevector}} &&
\vert{\bf q}\vert=q=k_1\sin\theta=k_0\sin\theta'=k_0\sqrt{1-\mu^2} &\cr
&&\ &&\ &\crr
&&\ &&\ &\cr
&&\hbox{azimuthal angle} &&\varphi &\cr
&&\ &&\ &\crr
&&\ &&\ &\cr
&&\matrix{{\rm reflection}\cr{\rm and}\cr{\rm transmission}\cr{\rm
coefficients}} &&\matrix{
\matrix{{\rm partial}\cr{\rm reflection}}:\left\{\matrix{
R=\left(\frc{P-p}{P+p}\right)^2
=\left(\frc{\mu-\sqrt{\mu^2-1+1/m^2}}{\mu+\sqrt{\mu^2-1+1/m^2}}\right)^2
\hfill\cr\vphantom{R}\cr
T=\frc{4Pp}{(P+p)^2}=\frc{4\mu\sqrt{\mu^2-1+1/m^2}}
{\left(\mu+\sqrt{\mu^2-1+1/m^2}\right)^2}\hfill\cr}\right.
\hfill\cr\vphantom{R}\cr
\matrix{{\rm total}\cr{\rm reflection}}:\left\{\matrix{
R=1\cr T=0}\right.\hfill} &\cr &&\ &&\ &\crr 
}}$$
\vfill\eject
\centerline {\bf Table 2}
\smallskip
$$\vbox{\init\halign to 16.5truecm
{\strut#&\vrule#\tabskip=1em plus 2em&
\hfil$#$\hfil&\vrule#&
\hfil$#$\hfil&\vrule#&
\hfil$#$\hfil&\vrule#&
\hfil$#$\hfil&\vrule#&
\hfil$#$\hfil&\vrule#&
\hfil$#$\hfil&\vrule#\tabskip 0pt\crr
&&\ &&\ &&\ &&\ &&\ &&\ &\cr
&&\ &&\tau_0 &&\tau_1(1)&&\gamma(1,1)&&B(0)&&\Delta\q&\cr
&&\ &&\ &&\ &&\ &&\ &&\ &\crr
&&\ &&\ &&\ &&\ &&\ &&\ &\cr
&&\hbox{isotropic}&&0.710446 &&5.036475 &&4.227681 &&1.881732 &&1/2 &\cr
&&\ &&\ &&\ &&\ &&\ &&\ &\crr
&&\ &&\ &&\ &&\ &&\ &&\ &\cr
&&\hbox{very anisotropic}
&&0.718211 &&5.138580 &&4.889703 &&2 &&0.555543 &\cr
&&\ &&\ &&\ &&\ &&\ &&\ &\crr
&&\ &&\ &&\ &&\ &&\ &&\ &\cr
&&\Delta(\%)&&1.1 &&2.0 &&15.7 &&6.3 &&11.1 &\cr
&&\ &&\ &&\ &&\ &&\ &&\ &\crr
}}$$
\bye